\documentclass[printer]{aa}
\usepackage{graphics,epsf,epsfig,graphicx}

\newcommand{\be}{\begin{equation}}
\newcommand{\ee}{\end{equation}}

\begin{document}

\title{The simulated $21\,\rm{cm}$ signal during the epoch of reionization~: full modeling of the Ly-$\alpha$ pumping  }

\author{S. Baek\inst{1}, P. Di Matteo\inst{1,2}, B. Semelin\inst{1}, F. Combes\inst{1}, Y. Revaz\inst{1,3}}
\offprints{sunghye.baek@obspm.fr}
\institute{
LERMA, Observatoire de Paris, UPMC, CNRS, 61 Av. de l'Observatoire, F-75014, Paris, France
\and 
GEPI, Observatoire de Paris, section de Meudon, CNRS, Place Jules Janssen, F-92190, Meudon, France
\and
Laboratoire d'Astrophysique, \'Ecole Polytechnique F\'ed\'erale de Lausanne (EPFL)
}
\date{Received ; accepted }
\authorrunning{Baek et al.}
\titlerunning{The simulated $21\,\rm{cm}$ signal during the Epoch of Reionization}

\abstract{The $21\,\rm{cm}$ emission of neutral hydrogen is the most promising probe of the epoch of reionization (EoR). In 
the next few years, the SKA pathfinders will provide statistical measurements of this signal, such as its power spectrum. 
Within one decade, SKA should produce a full tomography of the signal. Numerical simulations predicting these observations
are necessary to optimize the design of the instruments and help in the interpretation of the data. Simulations are 
reaching a reasonable level in terms of scale range, but still often rely on simplifications to compute the 
strength of the signal. The main difficulty is the computation of the spin temperature of neutral hydrogen which 
depends on the gas kinetic temperature and on the level of the local Lyman-$\alpha$ flux (The Wouthuysen-Field effect). A
$T_S \gg T_{CMB}$ assumption is usual. However, this assumption does not apply early in the reionization history, or even
later in the history  as long as the sources of X-rays are too weak to heat the intergalactic medium significantly. This work  presents the
first EoR numerical simulations including, beside dynamics and ionizing continuum radiative transfer, a self-consistent 
treatment of the Ly-$\alpha$ radiative transfer. This allows us to compute the spin temperature more accurately. We use two
different box sizes, 
$20\,h^{-1}\,\rm{Mpc}$ and $100\,h^{-1}\,\rm{Mpc}$, and a star source model. Using the redshift 
dependence of average quantities, maps, and power spectra, we quantify the effect of using different assumptions to 
compute the spin temperature and the influence of the box size.  The first effect comes from allowing
for a signal in absorption (i.e. not making the $T_S \gg T_{CMB}$ approximation).
The magnitude of this effect depends on the amount of heating 
by hydrodynamic shocks and X-rays in the intergalactic medium(IGM). With our source model we have little heating so regions seen in absorption
survive almost until the end of reionization.
The second effects comes from using the real, local, Lyman-$\alpha$ flux. This effect is important for an average ionization fraction of less than $\sim 10\%$~: it changes the overall amplitude of the $21\,\rm{cm}$ signal, and adds its own fluctuations to the power spectrum.
}

\maketitle

\section{Introduction}

The Epoch of Reionization (EoR) begins with the formation of the first sources of light resulting from the
non-linear growth of primordial density fluctuations. When this event occurs, which could be anytime
between $z\sim 30$ and $z\sim 15$, the intergalactic medium (IGM) is mostly cold, neutral and
optically thick at many wavelengths. Our main probe of baryonic
matter during this era is the $21\,\rm{cm}$ emission of neutral hydrogen, to which the IGM is
optically thin (Madau et al. \cite{Madau97}). To observe a $21\,\rm{cm}$ signal in 
emission or absorption against the cosmic microwave background (CMB), however, we need either
a sufficient local flux of Ly-$\alpha$ photons (Wouthuysen-Field effect, Wouthuysen \cite{Wouthuysen52},
Field \cite{Field58}) or a high enough baryon density (collisions), either of which will decouple
the hydrogen spin temperature from the CMB temperature. The first condition requires a sufficient 
number of sources and the second a sufficient growth of the density fluctuations. During the EoR, the Wouthuysen-Field effect is prevalent as it becomes efficient 
everywhere with time. Where the $21\,\rm{cm}$ signal turns on, its intensity is determined, in addition to the spin
temperature, 
by the neutral hydrogen density field which is the combination of the baryon 
density field and the complex distribution of growing ionized regions. This means that observing the $21\,\rm{cm}$
signal will teach us a lot about the reionization history. Presently, what is known about this history defines two constraints. The first is the value of $\tau$, the optical depth of the Thomson scattering of CMB photons by free electrons, which
is proportional to the column density of ionized hydrogen from us to the CMB.
This  value is $0.09 \pm 0.03$ for {\sl WMAP3} cosmology (Spergel et al. \cite{Spergel07}) and 
$0.087 \pm 0.017$ in {\sl WMAP5} cosmology (Komatsu et al. \cite{Komatsu08}).
In the over-simplified model of instantaneous reionization, this gives a reionization redshift 
of $z \sim 11$. In realistic models, $\tau$
provides a constraint on a redshift integral of the mean ionization fraction. The second constraint 
on reionization arises from
the spectra of high-$z$ quasars. The Gunn-Peterson trough appears as a sharp feature in the quasar spectra when the neutral 
fraction in the IGM surrounding the quasar rises above $1 \%$. From observations this happens at 
$z$ greater than $\sim 6$ (Fan et al. \cite{Fan06}). The reduced error bar on the value of $\tau$
in the {\sl WMAP5} results, together with the quasar constraint, strongly favor an extended reionization history. 

Much information is coded in the $21\,\rm{cm}$ emission. First, the nature of the sources
(Pop II, Pop III stars or  X-ray sources: QSO, SNe or X-ray binaries) and their formation history determine 
the evolution of the sky-averaged signal with redshift. It may also be 
possible to pinpoint specific events in the reionization history like an average ionization fraction of $0.5$ 
using the redshift dependence of the power spectrum (Mellema et al. \cite{Mellema06b}, Lidz 
et al. \cite{Lidz07b}). This type of information will be available in the next few years from
pathfinders like LOFAR or MWA. Within one decade, the Square Kilometer Array (SKA) will deliver a tomography (maps at many different redshifts) of the
signal at $1$ comoving Mpc resolution. If the different physics can be
separated in the signal (Barkana \& Loeb \cite{Barkana05a}, McQuinn et al. \cite{McQuinn06}), we will be 
able to extract information about the growth rate of large scale structures directly. If not,
it is possible to use them as independent source planes for gravitational
lensing by lower $z$ clusters and still deduce information on the structure growth
rate (Metcalf \& White \cite{Metcalf07}).

Numerical simulations of the $21$ cm signal are useful to explore the different possible source models and derive constraints for
the design of the future instruments in terms of frequency range, bandwidth, angular resolution and sensitivity. Comparison with
observations will then enable us to discriminate between source models and formation histories. Predicting the $21\,\rm{cm}$ signal requires
the knowledge of three local quantities~: the baryonic density field, the ionization fraction of hydrogen and the Ly-$\alpha$ flux 
(to compute the spin temperature). Moreover, high resolution is critical to obtain realistic, non-spherical ionized regions arising from 
the high photon consumption
in dense small scale structures, and large simulation boxes ($> 100\,h^{-1}\,\rm{Mpc}$) are necessary to correctly sample the large-scale clustering in the
source distribution and the abundance of rare sources (Barkana \& Loeb \cite{Barkana04}, Iliev et al. \cite{Iliev06}). With the development of 3D radiative transfer codes (Gnedin \& Ostriker \cite{Gnedin97}, Gnedin \& Abel \cite{Gnedin01}, 
Nakamoto et al. \cite{Nakamoto01}, Maselli et al. \cite{Maselli03}, Razoumov \& Cardall \cite{Razoumov05},
 Mellema et al. \cite{Mellema06a}, Rijkhorst et al. \cite{Rijkhorst06}, Susa \cite{Susa06}, 
Whalen \& Norman \cite{Wahlen06}, Trac \& Cen \cite{Trac07}, Pawlik \& Schaye \cite{Pawlik08}, Altay et al. \cite{Altay08})  and increases in computational power, it has become possible in the last few years to predict the $21\,\rm{cm}$ signal. But some 
compromises still have been necessary. The most common approximation is to assume $T_S \gg
T_{CMB}$, where $T_S$ is the hydrogen spin temperature and $T_{CMB}$ is the CMB temperature. 
In this regime the $21\,\rm{cm}$ brightness temperature
is independent of the spin temperature~: no information about the gas temperature and local Ly-$\alpha$ flux is needed. 
Obviously, this approximation fails if either the coupling of $T_S$ to $T_K$, the gas kinetic temperature, 
by Ly-$\alpha$ is weak (early and\slash or far from the sources), then $T_S \sim T_{CMB}$, or if the gas
is not significantly heated in the voids by X-rays. In other words, this approximation removes any possible absorption regime. However,
the signal seen in absorption has the potential to be stronger than in emission 
(Furlanetto et al. \cite{Furlanetto06a}).  Kuhlen et al. (\cite{Kuhlen06}), Shapiro et al. (\cite{Shapiro06}),
Gnedin \& Shaver (\cite{Gnedin04}) and Santos et al. (\cite{Santos07}) have probed the absorption regime in numerical simulations.  In Kuhlen et al. 
and Shapiro et al., this is done taking only the role of collisions into account in determining the $21\,\rm{cm}$ signal, i.e. ignoring the role of 
Ly-$\alpha$ pumping or prior to the existence of any Ly-$\alpha$ source.
Gnedin \& Shaver and Santos et al., in turn, add the effect of a Ly-$\alpha$  flux in their analysis, but  still
use a drastic simplification~: a homogeneous Ly-$\alpha$ flux changing with redshift. Considering a simple $r^{-2}$ profile around the sources to compute the Ly-$\alpha$ flux fluctuations,  Barkana \& Loeb (\cite{Barkana05b}) have shown that clustering and Poisson noise in the source distribution modify the $21\,\rm{cm}$ power spectrum. Furthermore, several recent works  (Semelin et al. \cite{Semelin07}, 
Chuzhoy \& Zheng \cite{Chuzhoy07}, Naoz \& Barkana \cite{Naoz08}) have shown that 3D radiative transfer effects modify the expected $r^{-2}$
dependence of the flux in a homogeneous medium, and even more so in an inhomogeneous density field.
The present work includes a full modeling of the
Ly-$\alpha$ radiative transfer and examines how the resulting $21\,\rm{cm}$ signal compares to the one using a homogeneous Ly-$\alpha$ flux.

Often working with the $T_S \gg T_{CMB}$ approximation, the early simulations used box sizes  equal to or smaller than 
$20\,h^{-1}\,\rm{Mpc}$ (Ciardi \& Madau \cite{Ciardi03},
 Gnedin \& Shaver \cite{Gnedin04}, Furlanetto et al. \cite{Furlanetto04}, Salvaterra et al. 
\cite{Salvaterra05}, Vald\'es et al. \cite{Valdes06}). Recent
works put the effort on simulating larger boxes, usually $\sim 100\,h^{-1}\,\rm{Mpc}$, 
while trying to preserve a good enough resolution (Mellema et al. \cite{Mellema06b}, 
Zahn et al. \cite{Zahn07}, Lidz et al. \cite{Lidz07a},
McQuinn et al. \cite{McQuinn07}, Lidz et al. \cite{Lidz07b}, Iliev et al. \cite{Iliev08}).
For the same resolution, radiative transfer simulations are usually more demanding than dark matter dynamical simulations. Consequently, a common strategy
is to run very high resolution DM simulations, derive the baryonic density field
with a constant bias, and compute a clumping factor within each larger scale radiative transfer cell. This is a method 
to include the effect of minihalos (Ciardi et al. \cite{Ciardi06}) as subgrid physics. In this method, the least robust
assumption is the use of a constant baryonic to DM density ratio~: while correct on a large scale, it fails
on the scale of clusters where hydrodynamics and heating\slash cooling processes play an important role.
In this work, we rely on self-consistent hydrodynamic simulations and do not use a clumping factor.

Section 2 presents the radiative transfer code used, Section 3 describes the simulation runs,
which are analyzed in Section 4. Section 5 summarizes the main results.

\section{The radiative transfer code~: LICORICE}

The LICORICE code integrates three main parts, a TreeSPH dynamical part which is not
used for this work (see section 3.1), and two radiative transfer parts for the ionizing continuum and the Lyman-$\alpha$
line. All three parts are parallelized using OpenMP.
  \subsection{Ionizing continuum radiative transfer}
We use a 3D Monte Carlo ray-tracing scheme. The radiation field is discretized
into photon packets which are propagated individually on an adaptive grid and interact with matter. 
The numerical methods implemented in LICORICE are similar to those presented by Maselli 
et al. (\cite{Maselli03}) for the code CRASH. 
There are, however, a number of differences which are described in the following subsections.

The version of LICORICE used in this work does not include H$_2$ formation, He ionization nor
diffuse radiation from recombination.

\subsubsection{Adaptive grid}

LICORICE uses an adaptive grid based on an oct-tree .
There is a difference between the usual oct-tree and our adaptive grid. 
The oct-tree keeps splitting a mother cell into 8 children 
cells if it contains more than 1 particle. However, the grid for RT stops  
splitting a mother cell if it contains less than a tunable parameter, $N_{max}$ particles. 
Therefore, the RT cells can contain from 0 to $N_{max}$ 
particles while the cells in the oct-tree contain either 0 or 1 particle.

For equivalent spatial resolution, such an adaptive grid requires lower memory and CPU-time  
than other grid-based RT codes. In this paper, we adopted $N_{max}=30$ for the continuum
ionization part, which makes the minimum RT cell size of 1 $h^{-1}$kpc at the end of the S20 simulation(z=5.6)
and 200 $h^{-1}$kpc for the S100 simulation(z=6.6)

    \subsubsection{SPH-particle density and grid-cell density}

We compute the gas density at each particle's position using the SPH smoothing kernel
(Monaghan \cite{Monaghan92}). We use on average 50 neighbor particles, $N_{neighbor}$,
to compute the SPH smoothing kernel.
We call it {\sl particle density} and consider that the
particle density is close to the physical density of the fluid (Hernquist \& Katz \cite{Hernquist89}).
We define another density: the average density of an RT cell that can
contain several particles.
We call it {\sl RT cell density} and use it to estimate the optical depth of each RT cell.
When a photon packet travels through an RT cell with optical depth $\Delta \tau$,
 the absorption probability is given by $P(\Delta \tau)=1-e^{-\Delta \tau}$.
For each RT cell crossed, we deposit a fraction of its initial photon content $\propto P(\Delta \tau)$. 
Then, we distribute the absorbed photons and energy to each particle in the RT cell proportionally
to its HI mass. We do not take into account any form of sub-cell density fluctuations. 
Indeed, particles in one RT cell have similar particle density because we limited
the maximum number of particles in one RT cell at $N_{max}=30$
which is smaller than $N_{neighbor}$, the number of neighbor particles of the SPH smoothing kernel.
Choosing a smaller $N_{max}$ requires using more photons, more memory and updating cells more frequently.
Updatings cells can be especially costly around the sources were cells recieve many photons.
We use the particle density to update physical quantities
such as the ionization fraction and temperature.
Therefore, we compute the photo-ionization rate
and temperature individually for each particle even if they are in the same RT cell.
Recombination and collisional ionization are also computed
individually for each particle.

This scheme is especially relevant if the density field is not static i.e. an
RT cell does not always contain the same set of particles.

   \subsubsection{Adaptive time integration}

\begin{figure}[t]
\centering
\resizebox{\hsize}{!}{\includegraphics{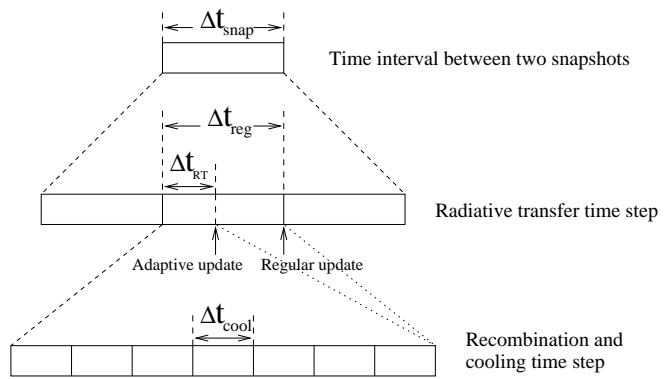}}
\caption{Schematic representation of the different time steps used in LICORICE. }
\label{diagram}
\end{figure}

The integration time step for updating the photoionization and photoheating rates, 
$\Delta t_{\rm{RT}}$, is adaptive. We use a typical $\Delta t_{\rm{RT}}$
much smaller than the time interval  $\Delta t_{\rm{snap}}$ between two snapshots
of the dynamical simulation.
Furthermore, recombination, collisional ionization and cooling are treated  
with an integration time step $\Delta t_{\rm{cool}}$ which is much smaller than
$\Delta t_{\rm{RT}}$. The photoionization and photoheating rates are kept constant
over all the $\Delta t_{\rm{cool}}$ in one $\Delta t_{\rm{RT}}$.
The relation between the different time steps is summarized  in Fig.~\ref{diagram}. 

The ionizing continuum radiative transfer is the repetition of
two steps. The emission and propagation of $N_{\rm{ph}}$ photon packets during 
the time interval $\Delta t_{\rm{reg}}$ (for continuum transfer an infinite 
velocity of light is assumed) and the update of physical quantities.

\begin{itemize}
\item i) Step 1~: All the sources emit a number $N_{\rm{ph}}$ of photon packets along
directions chosen at random. A photon packet leaves a fraction of its content of photon number
and energy in each RT cell it passes through. Each RT cell records 
the number of absorbed photons and energy during the emission of  the $N_{\rm{ph}}$
photon packets

\item ii) Step 2~: Then, the ionization state and the temperature are updated
with the integration time step $\Delta t_{\rm{reg}}$ for all particles in all RT cells.
This update is applied even to the RT cells that absorbed
no photon during $\Delta t_{\rm{reg}}$. This is necessary in order to consider the effect
of collisional ionization, recombination and adiabatic expansion of the gas (which is 
interpolated between snapshots).
In view of the $21\,\rm{cm}$ emission, the change of temperature due to the adiabatic
expansion of the Universe is especially important in diffuse regions,
independently of the photo-heating rate.
\end{itemize}

However, this regular update is not appropriate where the flux
of photons is too high (around sources for example). 
In this case, the number of absorbed photons can exceed the total number of atoms
available for ionization in the RT cell, or, in other words, the fixed update time step $\Delta t_{\rm{reg}}$
is much larger than the characteristic time scale of photo-ionization.
So, whenever the number of absorbed photons accumulated in a cell reaches a pre-set limit,
for example 30\% of the total number of neutral atoms in the cell, 
we update the physical quantities of the cell with the appropriate integration time step $\Delta t_{\rm{RT}} (< \Delta t_{\rm{reg}}) $.

When $N_{\rm{ph}}$ photon packets have been emitted, the integration time is synchronized 
as described in step 2 , using $\Delta t_{\rm{reg}}$ for the particles which were not 
updated since last synchronization and with the time elapsed since the last update 
for the others. The relative values of the different time steps are specified in section 3.2.1.

  \subsection{Ly-$\alpha$ line radiative transfer}

Computing Ly-$\alpha$ transfer is necessary to evaluate accurately the Wouthuysen-Field effect 
(see section 3.4) in the determination of the $21\,\rm{cm}$ brightness temperature.

The part of LICORICE which deals with the Ly-$\alpha$ line radiative transfer 
shares some features with its ionizing continuum counterpart~: mainly the Monte 
Carlo approach and the adaptive grid. For the Ly-$\alpha$ transfer, 
$N_{\rm{max}}=12$ is used. This is a smaller value than for the ionizing continuum
because Ly-$\alpha$ photons have no feedback on the gas: RT cells do not need to be updated
at all. There  is no direct CPU cost connected to using a smaller $N_{\rm{max}}$. 

The methods for resonant line 
scattering were presented in Semelin  et al. (\cite{Semelin07}). The 
implementation is similar to those of other existing codes (Zheng \& 
Miralda-Escud\'e \cite{Zheng02}; Cantalupo {\sl et al.} \cite{Cantalupo05}; 
Dijkstra {\sl et al.} \cite{Dijkstra06}; Verhamme {\sl et al.} \cite{Verhamme06}; 
Tasitsiomi \cite{Tasitsiomi06}). Necessary improvements have been 
implemented in our code for the simulations presented in this paper. Let us 
describe them briefly.
   \subsubsection{Taking the full effect of expansion into account}
Cosmological expansion is usually introduced in Ly-$\alpha$ transfer codes 
by including a Doppler shift associated with the Hubble flow velocities and 
by using a constant value for the expansion factor throughout the photon
flight. This works well for simulation boxes of moderate size (a few $10$
comoving Mpc). However, during reionization, a photon emitted just below 
Ly-$\beta$ will travel several hundreds of comoving Mpc before it redshifts to 
Ly-$\alpha$. Moreover, box sizes larger than $100$ comoving Mpc are necessary
to avoid underestimating the fluctuation power spectrum of the $21\,\rm{cm}$ line emission.
Consequently we took the necessary steps to include the full effect of 
cosmological expansion in our Ly-$\alpha$ simulations~:
\begin{itemize}
\item We use the comoving frequency. The usual approach is to use the 
frequency in the reference frame of the center of the simulation box and convert
back and forth to the frequency in the gas local rest frame by a Doppler shift
including peculiar and Hubble flow velocities. Instead we use the 
frequency in the local comoving frame as the reference frequency and include 
only peculiar velocities in Doppler shifts. If a photon enters a cell with 
local comoving frequency $\nu_{\rm{in}}$ at time $t_{\rm{in}}$ and exits the cell 
(or undergoes scattering) at time $t_{\rm{out}}$, its outgoing local comoving 
frequency $\nu_{out}$ is simply given by the general formula~:

\be 
\nu_{\rm{out}} = {a(t_{\rm{in}}) \over a(t_{\rm{out}}) } \nu_{\rm{in}} \quad,
\ee

\noindent
where $a(t)$ is the expansion factor of the universe.
As explained in Semelin et al. (\cite{Semelin07}), this avoids second 
order errors in ${\delta a \over a}$ where $\delta a$ is the variation of the 
expansion  factor during the flight of the photon through the {\sl whole} 
simulation box. However, between the incoming and outgoing comoving frequency 
in a cell, we do make a linear interpolation to compute the comoving frequency along the 
photon path. This is necessary to use Eq.~$13$ in Semelin et al. 
(\cite{Semelin07}) which allows a correct computation of the optical depth 
even when the photon redshifts all the way though the Ly-$\alpha$ line within the current cell.
The point is that in the current version, errors do not add up from cell to 
cell. 

\item The second benefit of using this formulation is that we track the variation
of the expansion factor $a(t)$ during the flight of the photon, and we can use
the correct value at any given time to compute such quantities as the gas 
physical (non-comoving) density. This avoids first order errors in ${\delta a \over a}$ when computing
the optical depth for example.

\end{itemize}

 In this framework, we naturally account for retarded time. The drawback is that we have to 
propagate over several box sizes (using periodic boundary conditions) some photons that were emitted more than 
10 snapshots earlier. This is necessary for the reasons mentioned above but also because the source population
changes drastically over such a time-scale.

    \subsubsection{Specific acceleration scheme}
A photon which redshifts through the Ly-$\alpha$ line in a hydrogen medium at 
the universe average density at $z \sim 10$ faces an optical depth of $\tau \sim 10^6$ 
 (Gunn \& Peterson \cite{Gunn65}, Loeb \& Rybicki \cite{Loeb99}). To compute $10^6$ scatterings 
for the $\sim 10^9$ photons
used in the simulations is impossible. An acceleration scheme has to be devised. The usual
method is the core-skipping scheme (Avery \& House \cite{Avery68}; Ahn et al. \cite{Ahn02}).
A photon entering the core of the line is systematically shifted out when it next scatters off a thermal atom.
This scheme reduces the number of scatterings and CPU cost by several orders of magnitude while 
preserving the shape of the emerging spectrum. However, in this work, we do not need to compute the
emerging spectrum. All we need is to know where the scatterings occur. Consequently we can use an even
more drastic acceleration scheme. Although the few scatterings that occur in the blue wing of the line
are important to determine in which location the photon enters the core of the line (Semelin et al. \cite{Semelin07}), their contribution to the total number of scatterings is negligible. Consequently, 
the following method is used~: we propagate the photon and compute scatterings until the local comoving
frequency is shifted within 10 thermal linewidths of the line center, in the blue wing, at a location 
$\mathbf{x_0}$. Then we consider that the $\sim 10^6$ scatterings that this photon is bound to 
undergo all occur at $\mathbf{x_0}$. At 10 thermal linewidths off the center, the mean free path of the 
photons is less than $1$ kpc and much shorter than $1$ pc in the core of the line. We can neglect 
the spatial diffusion while the photon remains within the 10 linewidth frequency range. 
As for the scatterings which occur outside the 10 linewidth range, 
there are just a few hundred of them and we can neglect their
contribution to the total number of scatterings (but not their importance in determining where the core
scatterings will occur). 

One factor remains to be specified~: exactly how this $\sim 10^6$ scattering 
number depends on the local state of the IGM. Looking at the optical depth computed for a redshifting
photon, three variable quantities can influence the number of scatterings~: the local neutral gas density, the local 
kinetic temperature of the gas and the current value of the Hubble constant. Escaping the
line is achieved by a combination of frequency shifts from two sources~: scattering off thermal atoms 
which produce a random walk in frequency, and the systematic shift to lower frequency between 
two scatterings due
to cosmological expansion. How these two effects combine is not clear at first sight~: although
cosmological shifts are systematically redward, they are much smaller than the random thermal shifts 
when the photon is still near the core of the line. We ran a series of simulations without any 
acceleration scheme at different values of the local parameters, for a single source in a homogeneous 
medium, and we computed the average number of scatterings per photon. It turns out that changing the 
temperature in a reasonable range (1 K to 1000 K) has no influence on the average number 
of scatterings (variation less than that the statistical error for 1000 photons).
The approach of Loeb \& Rybicki (\cite{Loeb99}), developed in a 0 K medium, remains valid in the 
cosmological context. This means, as was checked numerically, that the average number of scatterings
changes as $1/H(z)$. The number of scatterings per photon is obviously proportional to the overdensity
of neutral hydrogen,
however, in computing the $21\,\rm{cm}$ emission, we are interested in $P_{\alpha}$, the average number of 
scatterings per atom per second. In $P_{\alpha}$, the contribution of the overdensity cancels out. 
Consequently the following calibrated formula for $N_{\rm{scat}}$, the number 
of scatterings for each photon, was used~:

\be
N_{\rm{scat}}= 8\, \times 10^5 \,\,{H(z=10) \over H(z)}\,\,, 
\ee
where $H(z)$ is the Hubble constant at redshift $z$. Many non-linear
density structures at subgrid levels produce deviations from the above formula (Chuzhoy \& Shapiro \cite{Chuzhoy06}).
\section{Description of the simulations}

\subsection{Dynamics}

The dynamical simulations have been run using a modified version of the
parallel TreeSPH code Gadget2 (Springel \cite{Springel05}) which explicitly conserves
 both energy and entropy. 

\subsubsection{Initial conditions}

The initial conditions are the reference conditions of the Horizon Project\footnote{\texttt{http://www.projet-horizon.fr/}}. 
They have been computed using the public version Grafic-2 code (Bertschinger \cite{Bertschinger01}).
The cosmological parameters are those of the concordance
$\Lambda$CDM flat universe based on WMAP3 data alone (Spergel et al. \cite{Spergel07})~: 
$\Omega_{m}=0.24$, $\Omega_{\Lambda}=0.76$, $\Omega_{b}=0.042$,
with a normalization of the density fluctuation power spectrum given by $\sigma_8=0.76$.
The corresponding Hubble constant at the present epoch is 
$H_0=100\,h\,\rm{km}\,\rm{s}^{-1}\,\rm{Mpc}^{-1}$, with $h=0.73$.
 The initial redshifts ($z=42.7$ for the S20 and  $z=29.9$ for the S100 simulation)  were chosen 
using the default prescription from Grafic-2.
It relies on the Zel'dovich approximation to predict the growth of the smallest resolved structures until
they enter the non-linear regime. Then the simulations start. Crocce et al. (\cite{Crocce06}) show that 
this approach leaves some spurious transients compared to the more accurate second-order Lagrangian perturbation
theory. The corrections are of the order of a few percent on such quantities as the density power spectrum of the
dark matter halo mass function. We believe that, in EoR simulations, other uncertainties connected to the source
modeling outweigh this issue.

Using these cosmological parameters, we have run two simulations (S20 and S100) covering
different comoving volumes, respectively $20^3\,h^{-3}\,\rm{Mpc^3}$ and $100^3\,h^{-3}\,\rm{Mpc^3}$.
In both simulations, the number of particles is set to $2 \times 256^3$ ($\sim 33 \,\, \times 10^6$ particles),
where half corresponds to gas particles and half to dark matter particles.
The mass resolutions of dark matter and gas/stars particles, as well as the softening parameter, are given in tab.~\ref{simulation_parameters}.
 For comparison with other work, the minimum mass of resolved halos is $\sim 5\,\,\times 10^{10} M_{\odot}$ for the S100 and $\sim 4\,\,\times 10^8 M_{\odot}$ for the S20 
simulations. However, let us emphasize that we do not need to identify halos with a halo finder code: sources are formed self-consistently in the dynamical
code (see section 3.1.2).

\begin{table}
   \begin{tabular}{l c c c c}
\hline
\hline
    simulations   	& $L$			& $m_{\rm{DM}} $ 	& $m_{\rm{gas}} $ 	& $\epsilon$ \\	   
                	& $[h^{-1}\,\rm{Mpc}]$	& $[h^{-1}\,\rm{M}_{\odot}]$ 	& $[h^{-1}\,\rm{M}_{\odot}]$  	& $[h^{-1}\,\rm{kpc}]$\\
\hline
\hline
    S20			& $20$			& $2.6\times 10^{7}$	& $5.5\times 10^{6}$			& $2$	\\
    S100		& $100$			& $3.2\times 10^{9}$	& $6.9\times 10^{8}$			& $10$  \\

\hline
\end{tabular}
\caption[]{Mass and spatial resolution for the two simulations S20 and S100.}
\label{simulation_parameters}
\end{table}

\subsubsection{Gas physics}

Additional gas physics has been added to the public version of Gadget-2.
The cooling and heating of the gas are computed following the recipe
proposed by Katz et al. (\cite{Katz96}) where the gas is assumed to be optically thin
and in ionization equilibrium with the UV background (the latter corresponds to
updated values of the quasar radiation spectrum computed by Haardt \& Madau \cite{Haardt96}).
In addition to the heating and cooling rates given by Katz et al. (\cite{Katz96}),
we also have taken into account the inverse Compton cooling using the formula in Theuns et al. (\cite{Theuns98}).
These considerations apply to the dynamical simulation only, not to the radiative transfer simulations. 
Non equilibrium ionization
and heating-cooling processes are recomputed independently in the radiative transfer
simulations (see section 2.1).

Star formation is taken into account by transforming gas particles into star particles.
We have used the classical recipe that mimics a Schmidt law~:

	\begin{equation}
\frac{d \rho_\star}{dt} = \frac{\rho_g}{t_\star},
\end{equation}
\noindent
where $t_\star$ is defined by~:

\begin{equation}
t_\star= t_{0\star} \left( \frac{\rho}{\rho_{\rm{th}}} \right)^{-1/2}.
\end{equation}

Following Springel \& Hernquist (\cite{Springel03}), we have set $t_{0\star}=2.1\,\rm{Gyr}$ for the S20 simulation and a physical density
threshold of $\rho_{\rm{th}} =5.7\,\times 10^{-25}\,\emph{h}^2 \rm{g\,cm^{-1}}$.
Moreover, it is required that the local gas overdensity exceeds 200 in order to form stars.

Supernova feedback is taken into account by simply injecting in the surrounding gas particles
an amount of $10^{48}\,\rm{erg}$ per unit solar mass formed. To avoid  instantaneous dissipation by 
radiative cooling (Katz et al. \cite{Katz96}), $80 \%$ is ejected in kinetic form, and the rest
is ejected in thermal form.

\subsubsection{Calibration of the star formation history}\label{sfr_calibration}

To compare the $21\,\rm{cm}$ signals between the two simulations we 
need very similar star formation histories. However, the star formation rate non linearly depends on the numerical resolution, 
especially at high redshift (Springel \& Hernquist \cite{Springel03}, Rasera \& Teyssier \cite{Rasera06}). To avoid an extremely CPU time consuming fine tuning of $t_{0\star}$ in the
S100 simulation, we forced the star formation rate
to follow the one obtained in S20. This was performed by adapting at each
time step the parameter $t_{0\star}$, while relaxing the constraint on the minimum density needed to form stars.

\subsubsection{Output frequency}

Radiative transfer simulations are run as a post-treatment of the dynamical simulations, interpolating between the recorded snapshots.
To obtain a reliable interpolation, 125 snapshots were recorded from $z=41.66$ to $z=5.92$.
Conforming to the Horizon Project chosen rule, the interval between two snapshots is given in terms of the scale factor $a$ by~:

\begin{equation}
\Delta a = \frac{1}{2^{10}}.
\end{equation}

\subsection{Reionization}

\subsubsection{Initial conditions}

The kinetic temperature of the gas is computed in the dynamical simulations
assuming equilibrium for the ionization state with a uniform UV background.
However, it is recomputed in the radiative transfer simulations using the 
photo heating and cooling rate for a non equilibrium ionization state.
The radiative transfer for the S20 (resp. S100) simulation starts 
with the same initial temperature as the dynamical simulations, 
34.7K (resp.17.3K) computed with RECFAST (Seager et al. \cite{Seager99}).
Using a uniform temperature is an approximation: from the epoch of decoupling ($z \sim 150$),
different regions evolve adiabatically with different contraction/expansion factors
due to the growth of density fluctuations. Moreover, the decoupling from the CMB
is actually neither instantaneous nor homogeneous. Including these temperature fluctuations
 is not a simple task. We can estimate the rms amplitude of the fluctuations to be smaller than $10\%$.
We chose to ignore them.

The initial ionization fraction of gas is set equal to zero.
We used the recombination rate given by Spitzer (\cite{Spitzer78})
 and the collisional ionization in Cen (\cite{Cen92}).

Our requirement for choosing the value of the reionization time step $\Delta t_{\rm{reg}}$ is that the relative
variation of the gas temperature anywhere in the simulation box is less than $20 \%$.  This ensures that the recombination rate does not
change too fast within $\Delta t_{\rm{reg}}$. To meet this condition we used $\Delta t_{\rm{reg}}=1/40\,\,\Delta t_{\rm{snap}}$ (resp.$1/10\,\,\Delta t_{\rm{snap}}$) for S20 (resp. S100) and $\Delta t_{\rm{cool}}=1/100\,\,\Delta t_{\rm{reg}}$(both for S20 and S100).  Note that the S20 run required a smaller time step since it has a higher resolution. 

%We needed $1/40\,\,\Delta t_{\rm{snap}}$(resp.$1/10\,\,\Delta t_{\rm{snap}}$) for the 
%ionization timesteps($\Delta t_{\rm{reg}}$) for S20(resp. S100) simulations
%in order that the cooling could not decrease the gas temperature less than
%$20\%$ of its absolute value in one ionization timesteps. S20 required smaller
%time step since it has higher resolution.  

\begin{figure}[t]
\centering
\resizebox{\hsize}{!}{\includegraphics[angle=270]{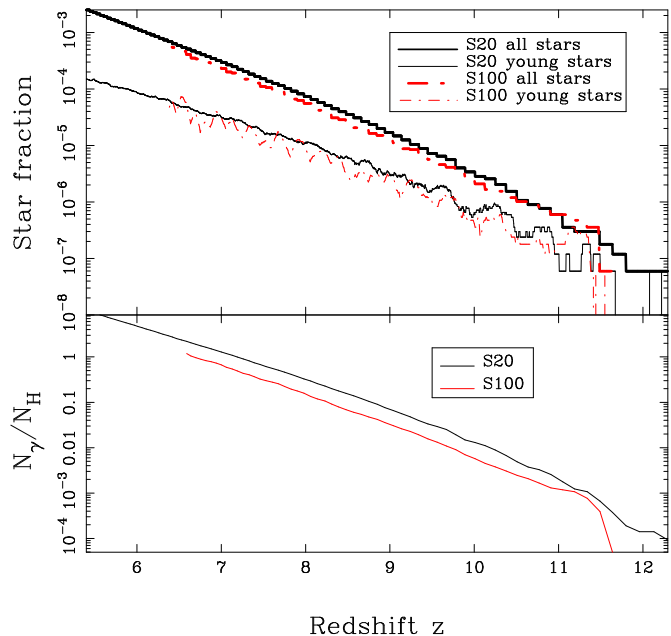}}
\caption{Top panel: Evolution of the fraction of baryonic mass locked into stars. Stars are considered young for the first 5-20 Myr. Bottom panel: Number of emitted photons per hydrogen atom.}
\label{star_fraction}
\end{figure}

\begin{figure}[t]
\centering
\resizebox{\hsize}{!}{\includegraphics{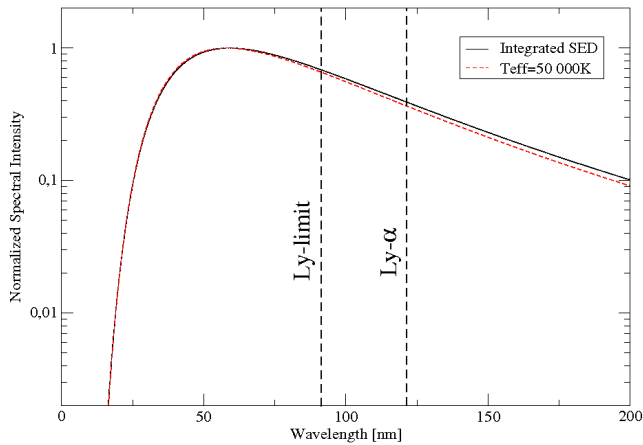}}
\caption{Normalized spectral intensity as a function of wavelength.}
\label{SED}
\end{figure}

\subsubsection{Post-treatment of the dynamical simulations: the issue of gas heating}
We use the snapshots of the dynamical runs as inputs for the radiative transfer simulation.
Consequently, there is no feedback on the dynamics from the radiative heating of the gas.
This has two opposite effects. First it slows down to some extent the propagation of the ionization 
fronts since the high gas pressure inside the Str\"omgren sphere is not effective. Second the source
formation is not quenched by radiative feedback~: stronger sources speed up the ionization front 
propagation. The net result of these two effects is unclear. In view of other important uncertainties,
like the source modeling, the feedback of radiative transfer on the dynamics is neglected.

The second consequence of running the radiative transfer simulation in post-treatment of the dynamical 
simulation is that dynamical effects on the temperature of the gas due to shock heating
are not included in the radiative transfer simulation.

 Shock heating is a sensitive issue since it directly affects the $21$ cm signal, possibly transforming
absorption into emission by raising the gas temperature above the CMB temperature. For example, mini-halos with
masses between $10^4$ to $10^8$ M$_{\odot} $ form very early during the EoR and are dense and warm enough from
shock heating during virialization to emit $21$ cm radiation. Iliev et al. (\cite{Iliev02}) and Shapiro et al. (\cite{Shapiro06})
show that the emission of mini-halos can dominate the $21$ cm signal prior to the onset of Ly-$\alpha$ coupling.
Furlanetto and Oh (\cite{Furlanetto06c}), however, find that the contribution of diffuse regions dominates later on.
The simulations presented here only marginally resolve the most massive of these mini-halos in the best
case (S20 simulation), and we chose not to try to include them as subgrid physics. But we should keep in mind 
that shock heating both below and above our scale resolution may affect the $21$ cm signal, as long as it occurs in
neutral regions (mini-halos, filaments).

The adiabatic cooling of the gas by dynamical expansion in low density regions is also an important 
factor responsible for the
possible strong absorption features in the $21\,\rm{cm}$ maps~: we can take it into account easily during the post treatment.
We use the fact that, below $T=10^4$ K (always true in neutral, $21\,\rm{cm}$ emitting regions), the cooling 
rate of the primordial gas is negligible, and shock heating is limited in the low density diffuse regions of the
IGM.
Consequently, we approximate the thermal evolution of the gas in the dynamical simulation as adiabatic.
We compute the variation of internal energy of a particle due to expansion or contraction
 between two consecutive snapshots from the variation of the density~:

\be
\bigtriangleup u = u_2 - u_1 = u_1\left( \left(\frac{\rho_2}{\rho_1}\right)^{\gamma -1} -1 \right),
\label{temperature}
\ee

\noindent
where $u$ is the internal energy and $\rho$ is the gas density. 

However, if the variation in density between two snapshots is large,
applying it all at once results in numerical instabilities. This happens when particles pass 
through or collapse  into a halo. To avoid this, the density is interpolated
gradually between two snapshots.

The other cooling and heating processes are then included self-consistently in the radiative transfer
simulation. We use the cooling-heating rates in Cen (\cite{Cen92}), and a minimum gas temperature of $1$K.

\subsubsection{Source modeling}

In this work we consider only stellar sources and do not study the influence of X-ray sources.
The decline of the quasar luminosity function at redshift $z>3$ suggests that stars were the dominant 
source of ionizing photons during the EoR (Shapiro \& Giroux \cite{Shapiro87}, or Faucher-Gigu\`ere et al. \cite{Faucher-Giguere08} for recent observational evidence).  However,
even in small quantities, X-ray sources have the potential to affect the $21$ cm signal by heating the IGM. 
Indeed X-ray photons can have a mean free path of several $100\,\rm{Mpc}$ through neutral hydrogen during the EoR: they
propagate through the ionization front out to the neutral regions. Using a homogeneous model Glover \& Brand 
(\cite{Glover03}) find that a modest number of X-ray sources can raise the temperature of the IGM by several
tens of K. Moreover, Pritchard \& Furlanetto (\cite{Pritchard07}) show that the heating by X-rays
is actually not homogenous: it produces fluctuations in the gas temperature which leave their own imprint on 
the $21$ cm power spectrum. In this work, however, we want to focus on the specific influence of Ly-$\alpha$ pumping.
X-ray sources will be included in a future work.

All baryon particles have the same mass~: $7.6\times10^6 M_{\odot}$ (resp. $9.5\times10^8 M_{\odot}$) for the
S20 (resp. S100) simulation.
Accordingly, one star particle corresponds in fact to a star cluster or a dwarf galaxy
so an IMF has to be assumed.
We choose a Salpeter IMF, with masses in the range 1M$_{\odot}$ - 100M$_{\odot}$. 
Then we compute the total luminosity  and spectral energy distribution (SED) for one star particle
by integrating over the IMF. We use the table
in Aller (\cite{Aller82}) to obtain the radius and
effective temperature of stars as a function of mass. The total ionizing luminosity (tab.~\ref{RT_parameters})
 of a star particle yields one or two photons per hydrogen atom at z=6.5 (Fig.~\ref{star_fraction}). 
The resulting SED is plotted in Fig.~\ref{SED}. It can be seen that it differs very little from a blackbody spectrum with 
$5\times10^4\begin{tiny}            \end{tiny}$ K effective temperature. Therefore our source model is roughly intermediate between PopII and PopIII stars.

\begin{table}
\centering
   \begin{tabular}{l c c }
\hline
\hline
    simulations   	& $L_{tot}[erg/s]$			& $f_{esc}$ 	\\
\hline
\hline
    S20			& $4.60\times 10^{43}$			& $0.08$	\\
    S100		& $5.75\times 10^{45}$			& $0.05$	 \\

\hline
\end{tabular}
\caption[]{The total ionizing luminosity of a star particle and escape fraction of simulations}
\label{RT_parameters}
\end{table}

  \subsubsection{Calibrating the global ionization history}
Before making a global calibration using the escape fraction, we need to consider the
effect of the time evolution of the stellar population within one source particle. Massive stars
contribute the most to the ionizing luminosity of the source, but live for a shorter time.
We use a simple model~: when a new source particle is created, we assign it a random lifetime  between 5Myr and 20Myr 
as a source of ionizing photons. The randomization is useful because we obtain new sources only with 
each new dynamical snapshot; a fixed lifetime for all sources and a discreetly evolving
number of sources with each new snapshot would produce distinct steps
in the average ionizing flux. Globally, the average lifetime of the source is degenerate with
the escape fraction as far as ionization history calibration is concerned, however, it allows
us to account for the signature of local starbursts (in time and space) followed by quiescence. 
We plot the effect of considering a finite lifetime for the ionizing source on the star fraction history in 
Fig.~\ref{star_fraction}. It can be seen that the young, ionizing star fraction fluctuates more strongly
at high redshifts than the global star fraction. A combination of local starbursts followed by star formation quenching 
and insufficient mass resolution may be responsible for this effect.

By trial and error, we calibrated the escape fraction (tab.~\ref{RT_parameters}) so that reionization is complete, at about $z \sim 6$.

\subsection{Lyman-$\alpha$ transfer}
We run the Lyman-$\alpha$ simulations as a further post-treatment of the reionization simulations.
The time interval between two snapshots is typically $10$ Myr at $z \sim 9$, during which the ionization
fronts travel a few hundred comoving kpc at most. We consider that this time interval is small enough
to permit a {\sl de facto} interpolation of the quantities relevant for Lyman-$\alpha$ transfer~: 
gas density and ionization state.

Potentially, Lyman 
series photons can heat up the gas (see references in Semelin et al. \cite{Semelin07}). 
But heating occurs only at low temperatures ($\sim 10$ K), where the gas pressure is insignificant.
Consequently, the feedback of Lyman-$\alpha$ photons on the dynamics is negligible.
However, in models with soft source spectra, where the efficient heating of low density 
neutral regions by X-ray photons is absent, one consequence of including the heating by 
Lyman-$\alpha$ photons is to increase moderately the temperature in the coolest regions. 
We implemented a local Lyman-$\alpha$ heating using the following formula, derived from Eq.~$47$ and Fig.~$3$ in 
Furlanetto \& Pritchard (\cite{Furlanetto06b})~:
\be
{dT_K \over dt}= {0.8 \over T_K^{1/3}} \,{x_\alpha \over S_\alpha} \,{10 \over 1+z} \,H(z)  \quad \mathrm{with} \quad S_\alpha \sim \left( 1 - {0.7 \over \sqrt{T_K}}\right)
\ee
We checked that when the $21\,\rm{cm}$ signal is in absorption with an intensity in the $300\,\rm{mK}$ range, i. e. where the gas 
temperature is a few K, this small amount of heating can reduce the signal intensity by up to $100\,\rm{mK}$.

Photons are emitted at frequencies between Ly-$\alpha$ and Ly-$\beta$. The spectrum and luminosity of
the  sources are computed with the method described in section 2.2.3 for the ionizing continuum. In principle
higher Lyman series photons also contribute. Indeed as soon as a photon emitted in the continuum hits an
atom in a resonant  upper
Lyman line,  it cascades locally down the levels and is transformed into a local Ly-$\alpha$ photon. These 
higher Lyman line photons are more concentrated around the sources since they need a smaller redshift 
before they resonate with a line (see
e.g. Naoz \& Barkana \cite{Naoz08}). Overall, including the upper Lyman line photons would increase the
number of Ly-$\alpha$ photons by a factor of less than 2. From the results of the simulations we estimate
that this could shift the history of the average Ly-$\alpha$ coupling by $\Delta z = 0.3$. It would also
add some limited amount of power in the fluctuations. In this work
we include only the Ly-$\alpha$ line. Upper Lyman series lines will be included in a future work.

In the simulations,  a fiducial number of $n_{\rm{ph}}=10^7$ photons is used between two dynamical snapshots. As the
time interval between snapshots and the total source intensity varies, the physical content of the photons
also varies. The value of $n_{\rm{ph}}=10^7$, dictated by the CPU cost of the simulations, is a lower limit for 
statistical significance~: indeed, with our current grid resolution of less than 12 particles per cell, 
about $10 \% $ of the cells do not receive any photons between two snapshots. These cells are 
small and are mostly located in moderate density filaments and low flux environments, far from the sources.
They are properly averaged with some of their neighbors when the fixed-grid maps are computed. To assess
the impact of this limited sampling we also ran a simulation with $6 \,\times10^7$ photons between two 
snapshots down to $z=9.5$ for the $20$ h$^{-1}$~Mpc box. In this case the fraction of unsampled 
cells drops to $\sim 1 \%$. It is only when  maps are drawn with thin slices, $\sim 200\,h^{-1}\,\rm{kpc}$,  
that a difference can be seen. 

\subsection{Physics of the $21\,\rm{cm}$ emission}

To compute the differential brightness temperature of the $21\,\rm{cm}$ emission in the optically thin limit of the $21\,\rm{cm}$ line,
we use the formula~:

\begin{eqnarray}
\delta T_b = 28.1 \,\,{\mathrm{mK}}\,\,x_{\mathrm{HI}}\,\, (1+\delta) \left({1+z \over 10}\right)^{1 \over 2}\,\, {T_S -T_{\mathrm{CMB}} \over T_S}  \nonumber\\
\hspace{1.5cm} \times \left( {\Omega_b \over 0.042} {h \over 0.73} \right) \left(0.24 \over \Omega_m \right)^{1 \over 2} \left( {1-Y_p \over 1-0.248} \right)  \,, 
\end{eqnarray}

\noindent
where $\delta$ is the local overdensity at redshift $z$, $x_{\mathrm{HI}}$ the neutral
hydrogen fraction, $T_S$ the neutral hydrogen spin temperature, $T_{\mathrm{CMB}}$ the CMB radiation 
blackbody temperature at redshift $z$ and $\Omega_b$, $\Omega_m$, $h$ and $Y_p$ are the usual 
cosmological parameters. In this formula, the contribution from the proper velocity
gradients have been neglected (Barkana \& Loeb \cite{Barkana05a}). 
 Mellema et al. (\cite{Mellema06b}) include this effect in computing $21$ cm line of sight spectra.
Analytic (McQuinn et al. \cite{McQuinn06}) and 
semi-numerical models (Mesinger \& Furlanetto \cite{Mesinger07})  find that the contribution of proper 
velocity gradients to the $21$ cm power spectrum is important
only at $\langle x_{\mathrm{HI}} \rangle < 0.5$ on scales smaller than the effective ionization 
bubble size. Since previous simulations did not include a proper treatment of the Ly-$\alpha$ pumping
which is especially important during the early phase of reionization, it was not relevant to include
the effect of proper velocity gradients. In this work, we include a full modeling of the Ly-$\alpha$ 
pumping. In a separate paper, the effect of proper velocity gradients will be studied.

The spin temperature of hydrogen $T_S$ is coupled to the CMB temperature $T_{\mathrm{CMB}}$ by 
Thomson scattering of CMB photons, and to the kinetic temperature of the gas $T_K$ by collisions 
and Ly-$\alpha$ pumping. As a result, the spin temperature can be written
(Furlanetto et al, 2006a)~:

\be
T_S^{-1}={ T_{\mathrm{CMB}}^{-1} +x_c T_K^{-1} + x_{\alpha} T_C^{-1} \over 1+x_c+x_{\alpha}} \quad \mathrm{with} \quad T_C \simeq T_K \,,
\ee
\noindent
and
\be 
x_{\alpha}={4 P_{\alpha} T_{\star} \over 27 A_{10} T_{\mathrm{CMB}} }, \quad   
x_c={T_\star \over A_{10} T_{\mathrm{CMB}}}(C_H+C_p+C_n), \,
\ee 
\noindent
where $T_{\star}={h\nu_0 \over k_B}= 0.068\,\rm{K}$ ($\nu_0=1420\,\rm{MHz}$), $A_{10}= 2.85 \times 10^{-15} s^{-1}$ is the spontaneous emission
factor of the $21\,\rm{cm}$ transition, $P_{\alpha}$ is the number of scatterings of Ly-$\alpha$
photons per atom per second, and $C_H$, $C_p$ and $C_e$ are the de-excitation rates due to collisions
with neutral atoms, protons and electrons. For the de-excitation rates we use the fitting formula given
by Liszt (\cite{Liszt01}) and Kuhlen et al. (\cite{Kuhlen06}). The
coupling coefficients $x_\alpha$ and $x_c$ are computed for each gas particle from the simulation data.
The final output is the value of the brightness temperature for each particle in each snapshot.

\begin{figure}[t]
\centering
\resizebox{\hsize}{!}{\includegraphics{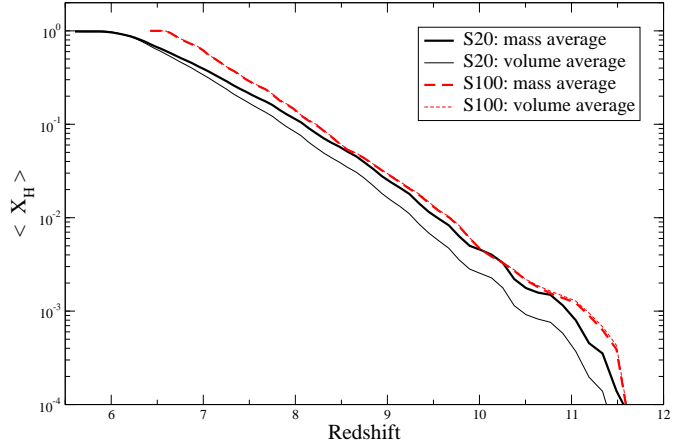}}
\caption{ Mass weighted and volume weighted averaged ionization fraction.}
\label{xH_avg}
\end{figure}

\section{Analysis}
We will focus our analysis on two aspects.

 First, by comparing our two simulations, we will 
evaluate the effect of numerical resolution and box size. Numerical resolution truncates the physics
at small scales~: this mainly modifies the reionization history and geometry by removing small, dense 
structures with high recombination rates. The limited box size, on the other hand, truncates large
scales and dampens the  $21\,\rm{cm}$ brightness temperature power spectrum on corresponding wavenumbers.

The second axis of our analysis is to measure the effect of including an accurate treatment of the 
Wouthuysen-Field effect in computing the brightness temperature.

\begin{figure}[t]
\centering
\resizebox{\hsize}{!}{\includegraphics{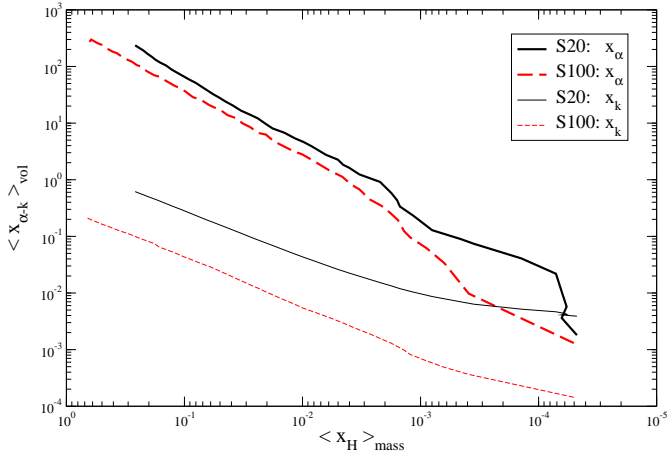}}
\caption{Evolution of the volume averaged Lyman-$\alpha$ and collisional coupling coefficients as functions of the average
ionization fraction for both simulations (see main text for a definition of the coefficients).}
\label{xa_xk}
\end{figure}

  \subsection{Evolution of sky-averaged quantities}
\subsubsection{Ionization fraction}
The simplest way to characterize the global history of reionization is to compute the average 
ionization fraction as a function of redshift. The mass weighted ($\langle x_H \rangle_{\rm{mass}}$) and volume weighted ($\langle x_H \rangle_{\rm{vol}}$) average ionization
fractions are shown in Fig.~\ref{xH_avg} for both simulations. For the volume (resp. mass) average we use
the volume occupied by each particle derived from the SPH smoothing length (resp. the mass of each particle)
as weights in the averaging procedure. The main feature of this plot is that
the mass weighted averages are similar for the two simulations while the volume weighted averages
differ. Indeed, the star formation histories of the two simulations were calibrated to produce the same
number of ionizing photons at any redshift. As long as recombination is negligible, this should produce
identical mass weighted ionization fractions. It can be checked in Fig.~\ref{xH_avg} that below redshift
$\sim 8$ recombination becomes efficient in the S20 simulation, which contains denser small-scale 
structures due to higher resolution, and the mass weighted ionization fraction drops below its S100 
counterpart. 

The volume weighted ionization fractions differ at all redshifts between the two simulations,
because highly ionized regions of identical mass, close to the sources,  are denser and occupy a 
smaller volume in the more resolved S20 simulation. McQuinn et al. (\cite{McQuinn06}) advocate the use of
the volume-averaged ionization fraction as an evolution variable instead of redshift. Indeed, depending
on arbitrary choices (all consistent with current observational constraints) for the source formation
history and source nature, reionization can begin at $z=12$, $15$ or $20$ and proceed differently. However,
for a fixed average ionization fraction, McQuinn et al. (\cite{McQuinn06}) show that the brightness 
temperature power spectrum depends weakly on the redshift at which this ionization fraction is achieved.
This is true if a modification of the source formation efficiency is the main cause of the change in redshift 
for a fixed ionization fraction. If the change is caused by using a different source spectrum, or modifying the gas
sub-grid physics (e.g. an inhomogenous clumping factor), this may not be true anymore. We believe
that the averaged ionization factor is indeed a more relevant measure of the evolution of reionization than redshift 
and we adopt their point of view. However, instead of the volume averaged ionization fraction, which
is well suited to comparing different models with the same underlying gas density field, we will use
the mass averaged ionization fraction which is better suited for the comparison of simulations with 
different mass resolution. While the mass averaged ionization fraction is a suitable quantity 
to track the progression of reionization, a volume average is more relevant to compute such observable
quantities as the $21\,\rm{cm}$ brightness temperature. Consequently, all quantities other than the ionization
fraction will be volume averaged.

We also computed the Thomson optical depth from the two simulations ignoring the presence of helium.
The values are $\tau=0.060$ and $\tau=0.063$ for S20 and S100 simulations respectively, i.e. within $1\sigma$ of the WMAP3 value $\tau=0.0089\pm0.030$ (see Spergel et al.  \cite{Spergel07}).

\begin{figure}[t]
\centering
\resizebox{\hsize}{!}{\includegraphics[angle=-90]{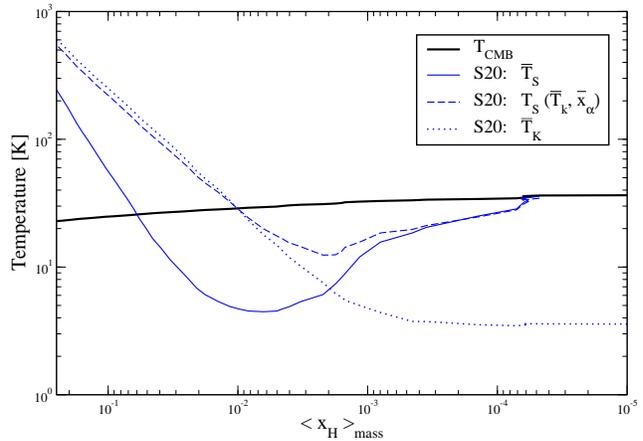}}
\caption{Evolution of the neutral hydrogen spin temperature as a function of the average ionization fraction. The average
is computed either directly from the particle spin temperatures or from the volume averaged values of $T_K$ and 
$x_\alpha$. The CMB temperature and volume averaged kinetic temperature are plotted for comparison. }
\label{Ts_two_avg_fn_xhmass}
\end{figure}

\subsubsection{Spin temperature coupling coefficients}

The spin temperature of neutral hydrogen, $T_S$, is defined by Eq.~$7$ and depends on two coupling
coefficients, $x_\alpha$ for the Wouthuysen-Field coupling and $x_k$ for collisional coupling. Until
now, $x_\alpha$ has never been computed consistently in simulations of the $21\,\rm{cm}$ emission. While
$x_\alpha$ itself contains fluctuations which contribute to the  power spectrum of the  $21\,\rm{cm}$ signal,
we study first the evolution of the average value of the coefficients which  tells us at which
stage in the reionization history the spin temperature can be considered fully coupled to the kinetic
temperature of the gas. The evolution of the volume averaged coefficients $\langle x_\alpha\rangle_{\rm {vol}}$ and $\langle x_k\rangle_{\rm{vol}}$ is shown in Fig.~\ref{xa_xk}, as
a function of  $\langle x_H \rangle_{\rm{mass}}$. 
The first conclusion from Fig.~\ref{xa_xk} is that $x_k$ is negligible compared to
$x_\alpha$ except in the very early phase ($\langle x_H \rangle_{\rm{mass}} \sim 10^{-4}$) for the S20 simulation. Logically,
the S20 simulation, producing denser structures, has a higher $x_k$ (indeed $C_H$, $C_p$ and $C_e$ in
Eq.~$8$ depend on the density).  With a better mass resolution and even denser structures than in the S20 simulation, $x_k$ would
still increase, and the ionization fraction when $x_\alpha$ starts to dominate would also increase.
The second piece of information that can be derived is the magnitude of the
error made by applying the usual assumption $x_\alpha=+\infty$. When $x_\alpha >10$, the error
made on $\delta T_b$ by assuming $x_\alpha=+\infty$ is smaller than $\sim 10 \%$. This corresponds
to $\langle x_H \rangle_{\rm{mass}} > 0.02$ for the S20 simulation and $\langle x_H \rangle_{\rm{mass}} > 0.04$ for the S100 
simulation. It may appear that the full coupling occurs very early on. Let us emphasize, however, that
in this early phase there is a strong absorption signal (with our choice of source model) compared
with the weaker emission signal observed later on. In a source model with a harder spectrum
(to be studied in a future paper) we would have a weaker absorption signal. If the sources are still stars 
( e.g. using a more top-heavy IMF) the full coupling would
also occur at a larger average ionization fraction since the ratio of ionizing to Lyman-alpha photons
would be larger.

\begin{figure}[t]
\centering
\resizebox{\hsize}{!}{\includegraphics[angle=-90]{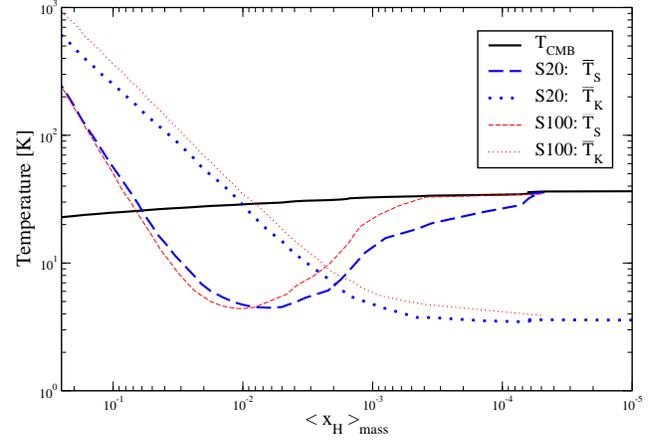}}
\caption{The evolution of the volume averaged spin temperature, kinetic temperature, and CMB temperature is plotted as
a function of the average ionization fraction. The quantities are plotted for both box sizes. }
\label{Ts_20_100_fn_xhmass}
\end{figure}

\subsubsection{Spin temperature}

 Fig.~\ref{Ts_two_avg_fn_xhmass} displays the average spin temperature for the S20 simulation, with the average computed 
in two different ways. First the spin temperature is computed using the volume averaged quantities  $\langle T_k\rangle_{\rm{vol}}$,  $\langle x_\alpha\rangle_{\rm{vol}}$ (and  $\langle x_k\rangle_{\rm{vol}}$), following Furlanetto et al. (\cite{Furlanetto06a}) and Gnedin \& Shaver (\cite{Gnedin04}).
This quantity is denoted $T_S(\bar{T}_K,\bar{x}_\alpha)$. Then the more relevant value, $\bar{T}_S$, is computed, as the direct volume average on
the particles of the spin temperature, i.e. $\bar{T}_S=\langle T_S\rangle_{\rm{vol}}$. Fig.~\ref{Ts_two_avg_fn_xhmass} shows that there is a large difference
between the two definitions. This is mainly due to the fact that the spin temperature is computed through the weighted average of
{\sl inverse} temperatures (Eq.~$7$). This large difference propagates to the average brightness temperature. Consequently
we strongly advocate replacing the Furlanetto et al. average traditionally used to define the emission and absorption
regimes by the more relevant direct average of the spin/brightness temperature. Of course this is easily done only 
for simulations. Let us mention that with the direct average, the spin temperature remains below the CMB temperature
at lower $z$.

Fig.~\ref{Ts_20_100_fn_xhmass}  compares the direct average of the particle spin temperature in the 
two simulations~: the
results are quite similar. The main difference is that the spin temperature starts to decouple from the CMB temperature 
later in the S100 simulation. Indeed, we use periodic boundary conditions in the simulations. Thus the Lyman-$\alpha$
sources are more homogeneously distributed in the $20\,h^{-1}\,\rm{Mpc}$ box, which lacks large scale modes, than in 
the $100\,h^{-1}\,\rm{Mpc}$ box (Barkana \& Loeb \cite{Barkana05b}). In the $100\,h^{-1}\,\rm{Mpc}$ box, large void regions 
receive little Lyman-$\alpha$ flux. There, the spin temperature remains coupled to the CMB at lower redshifts.

\begin{figure}[t]
\centering
\resizebox{\hsize}{!}{\includegraphics[angle=-90]{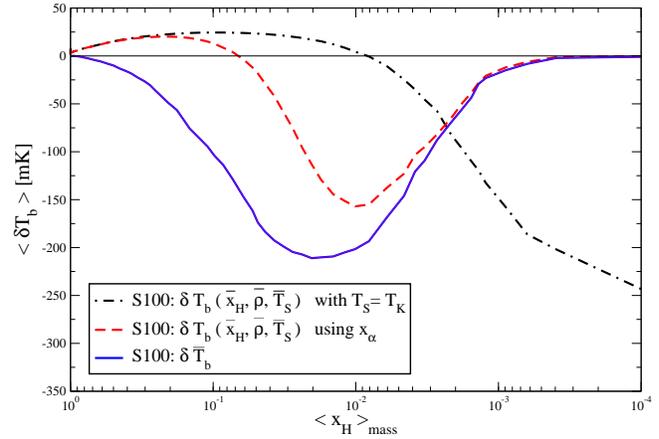}}
\caption{ Evolution of the average differential brightness temperature for the S100 simulation. The average is computed
in three different ways (see main text and legend). The blue curve is the most relevant direct average of the particle brightness
temperature. }
\label{Tb_two_avg_fn_xhmass}
\end{figure}

\subsubsection{Differential brightness temperature}

\begin{figure}[t]
\centering
\resizebox{\hsize}{!}{\includegraphics[angle=-90]{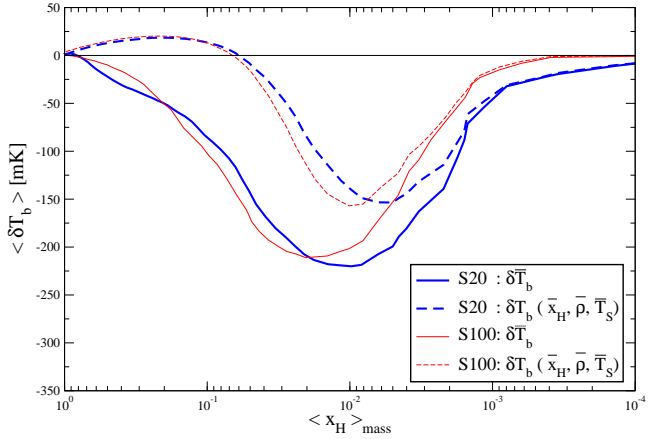}}
\caption{Comparison of the evolution of the average differential brightness temperature for the two box sizes. The average
is computed in two different ways (see main text). }
\label{Tb_20_100Mpc_fn_xhmass}
\end{figure}

The differential brightness temperature is plotted in Fig.~\ref{Tb_two_avg_fn_xhmass} for the S100 simulation. The 
average is computed in three different ways~: $\delta T_b(\bar{x}_H,\bar{\rho},\bar{T}_S)$ with $T_S=T_K$ for each particle, 
$\delta T_b(\bar{x}_H,\bar{\rho},\bar{T}_S)$ computing $T_S$ for each particle using $x_\alpha$, and $\bar{\delta T}_b$, 
using the same notations as for the spin temperature. In all these definitions we use a volume weighted average.
The first definition uses the common $T_S=T_K$ assumption~: as can be seen in the plot, it 
fails at early times when the coupling is weak. The second definition (equivalent to Furlanetto et al. 
\cite{Furlanetto06a}) shows an emission and an absorption regime. However, with the more realistic third definition,
 $\bar{\delta T}_b$, the emission regime disappears completely. 
Indeed, absorption region have a strong signal which dominated over the weak, $\sim 25\,\rm{mK}$ signal of the emitting 
regions. Of course, if X-ray sources were included, the cold neutral IGM would be pre-heated
and the absorption regions contribution would shrink much faster~:  $\bar{\delta T}_b$ would show an
emission regime and the absorption regime would be weaker. Including shock heating would go in the same direction although
the magnitude of the change would be less homogenous and is more difficult to estimate.

To assess the effect of resolution and box size, we plot in Fig.~\ref{Tb_20_100Mpc_fn_xhmass} the direct $\bar{\delta T}_b$ and $\delta T_b(\bar{x}_H,\bar{\rho},\bar{T}_S)$ for both simulations. The overall behaviour is similar. In the S20 
simulation the minimum of $\bar{\delta T}_b$ occurs at a $1\%$ average ionization fraction compared to a $2\%$ ionization 
fraction for the S100 simulation. This small difference is due in part to the slightly smaller coupling of the S100 
simulation at equal ionization fraction. The values of the minimum $\bar{\delta T}_b$ are not very different either. We 
expect to find resolution effects and box size effects in the power spectra.

We have presented these average values to compare  with previous studies. However, much information
is erased when considering average quantities, which makes their interpretation of limited scope. We proceed now to
present maps of the different quantities.

  \subsection{Maps}

\subsubsection{Ionization fraction}

\begin{figure*}[p]
\centering
\begin{array}[t]{ccc}
\resizebox{7.5cm}{!}{\includegraphics[angle=270]{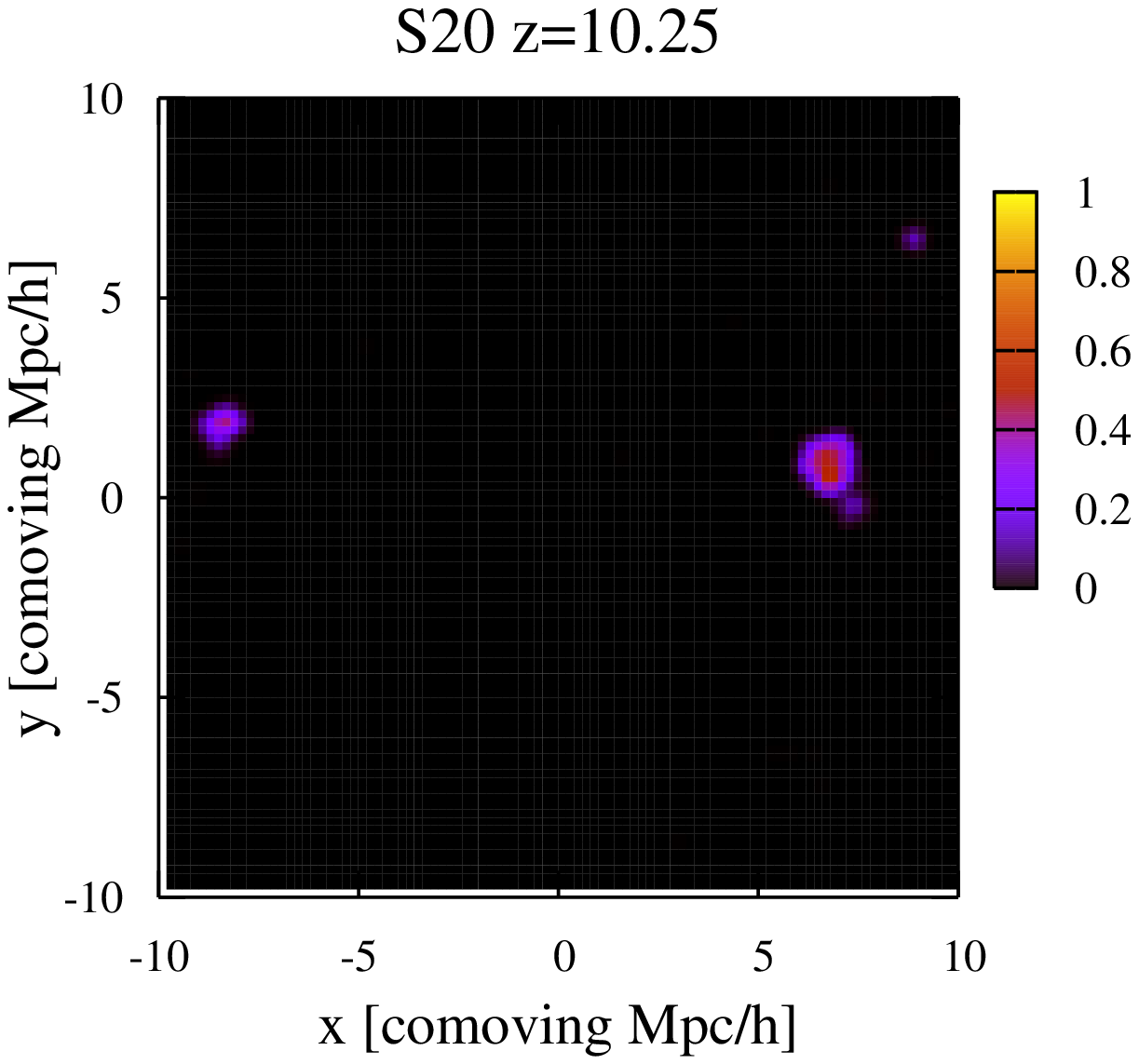}} &
\hskip 1cm\rotatebox{90}{\hskip -5.5cm \hbox{\bf \large $2 \,\,\,\mathrm{Mpc/h}$ (comoving) slice} }\hskip 1cm&
\resizebox{7.5cm}{!}{\includegraphics[angle=270]{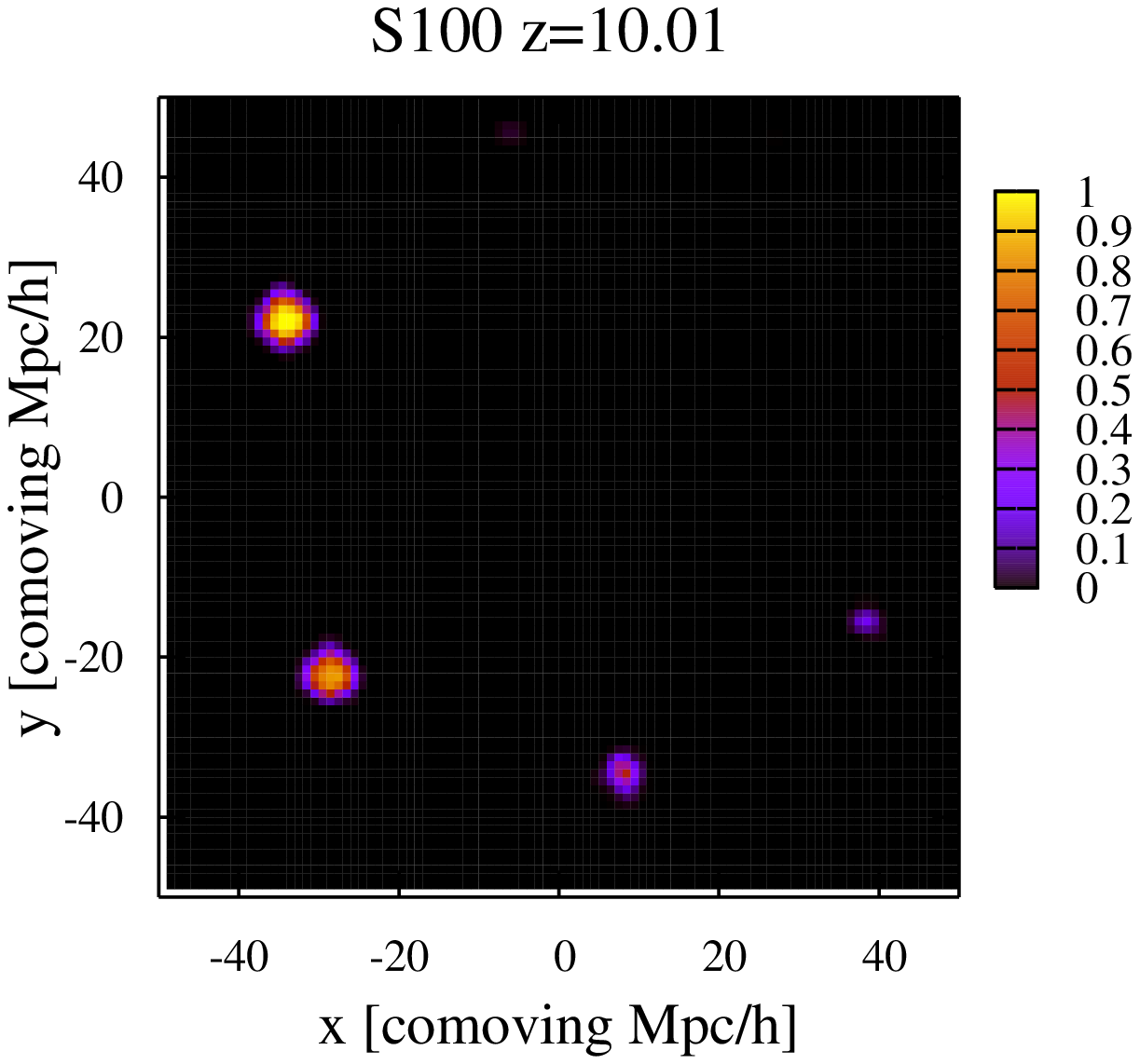}} \\
\hline
\resizebox{7.5cm}{!}{\includegraphics[angle=270]{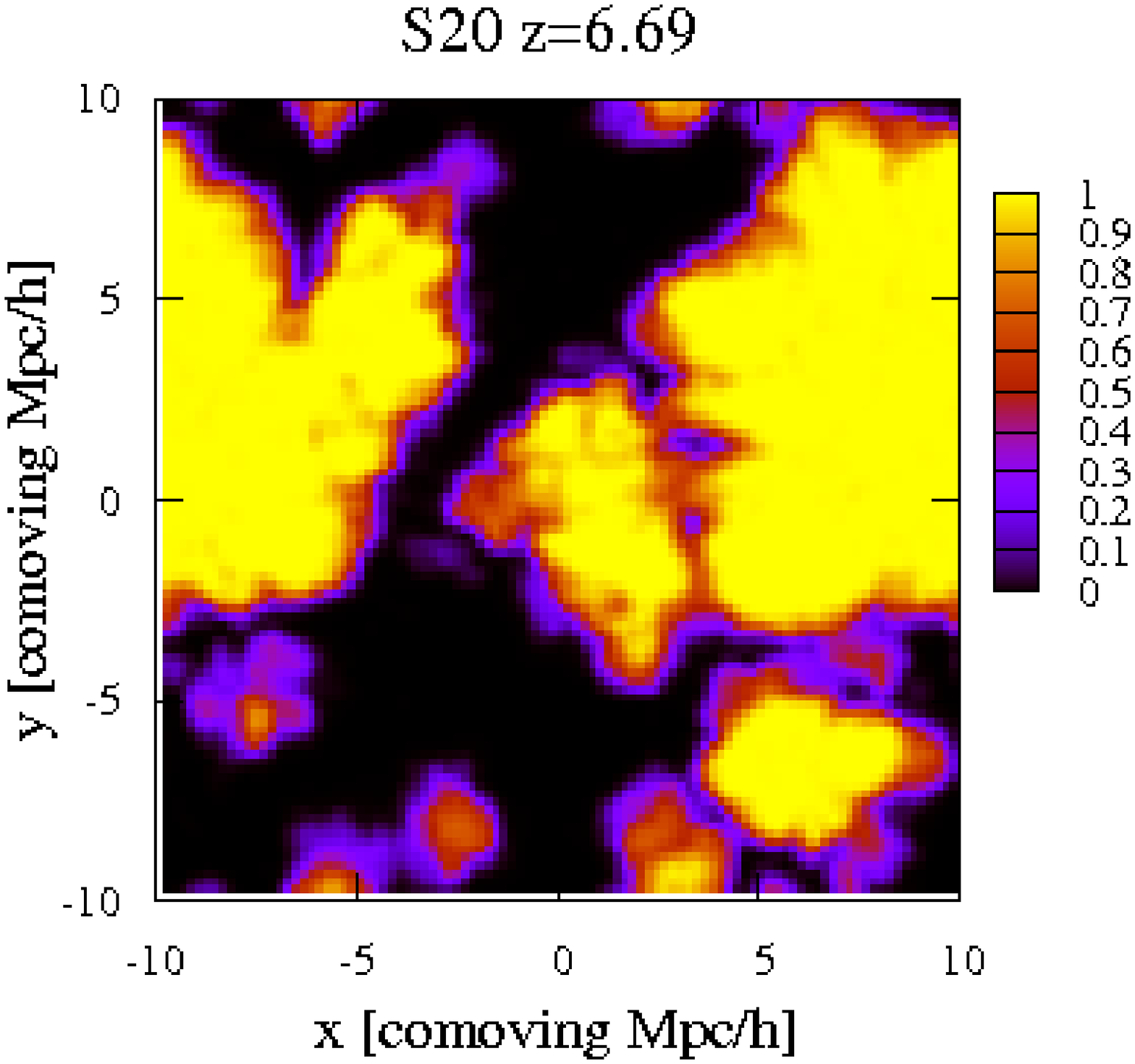}} &
\hskip 1cm\rotatebox{90}{\hskip -5.5cm \hbox{\bf \large $2 \,\,\,\mathrm{Mpc/h}$ (comoving) slice} }\hskip 1cm&
\resizebox{7.5cm}{!}{\includegraphics[angle=270]{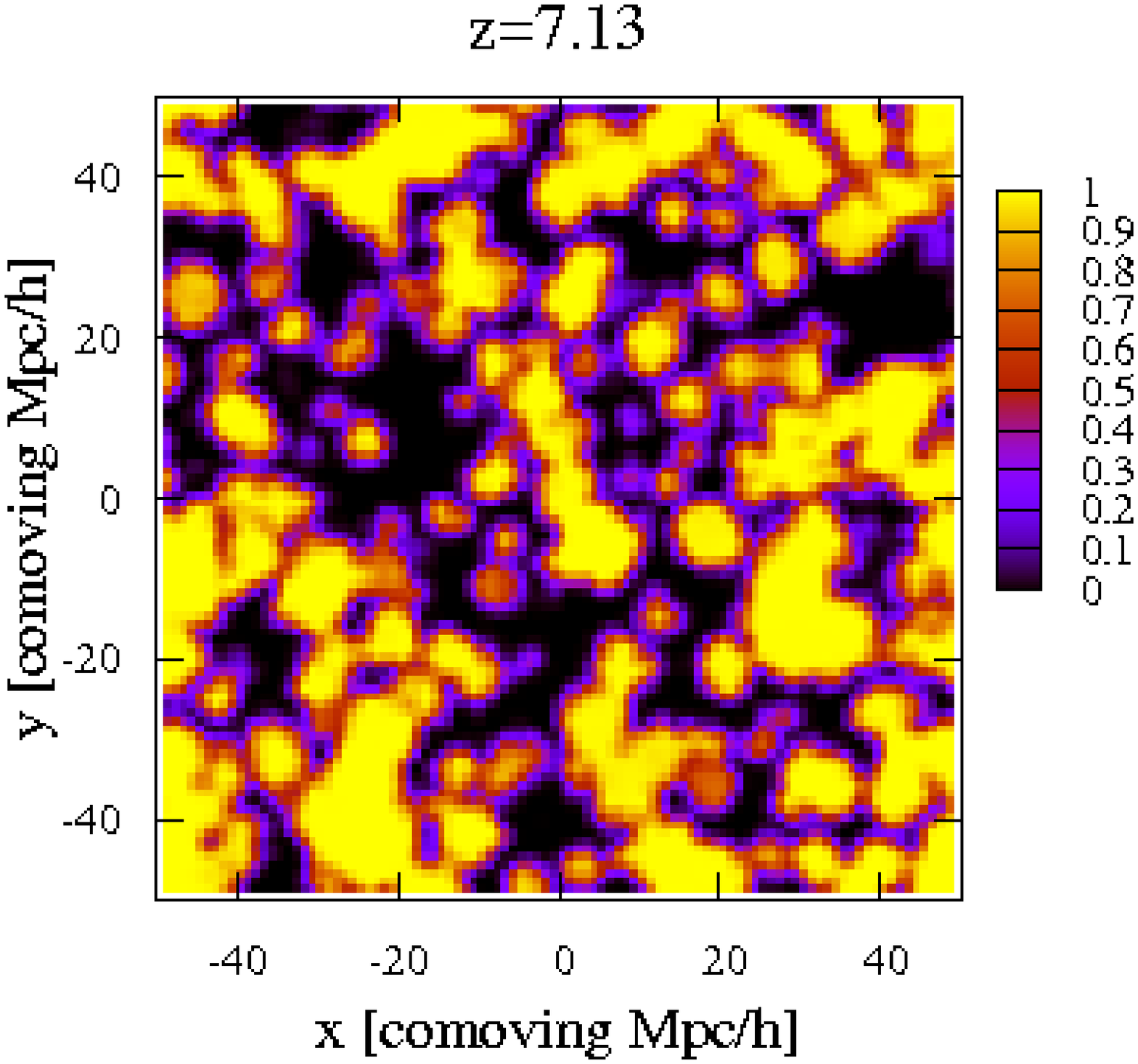}} \\
\resizebox{7.5cm}{!}{\includegraphics[angle=270]{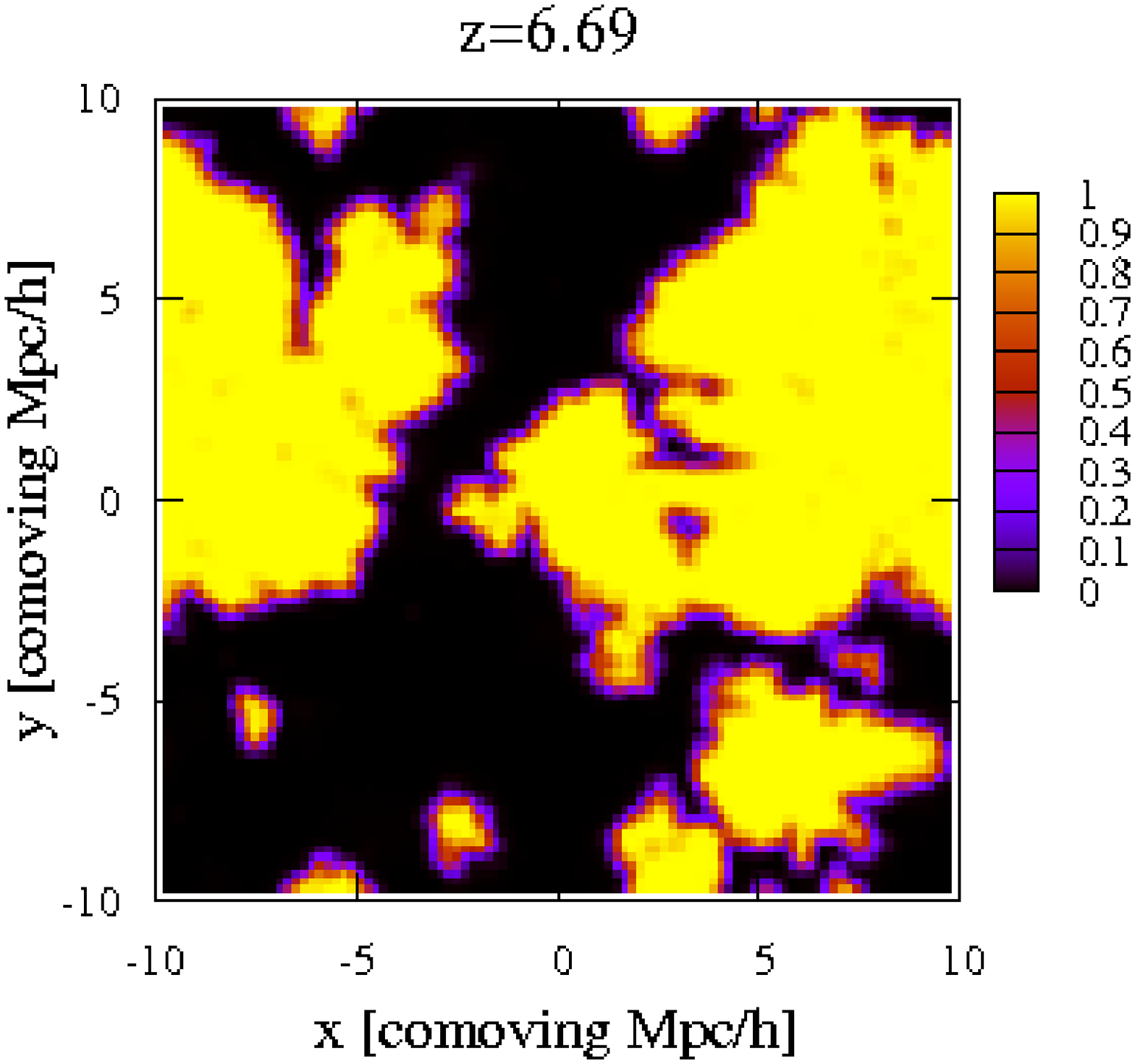}} &
\hskip 1cm\rotatebox{90}{\hskip -5.5cm \hbox{\bf  \large $200 \,\,\,\mathrm{kpc/h}$ (comoving) slice} }\hskip 1cm&
\resizebox{7.5cm}{!}{\includegraphics[angle=270]{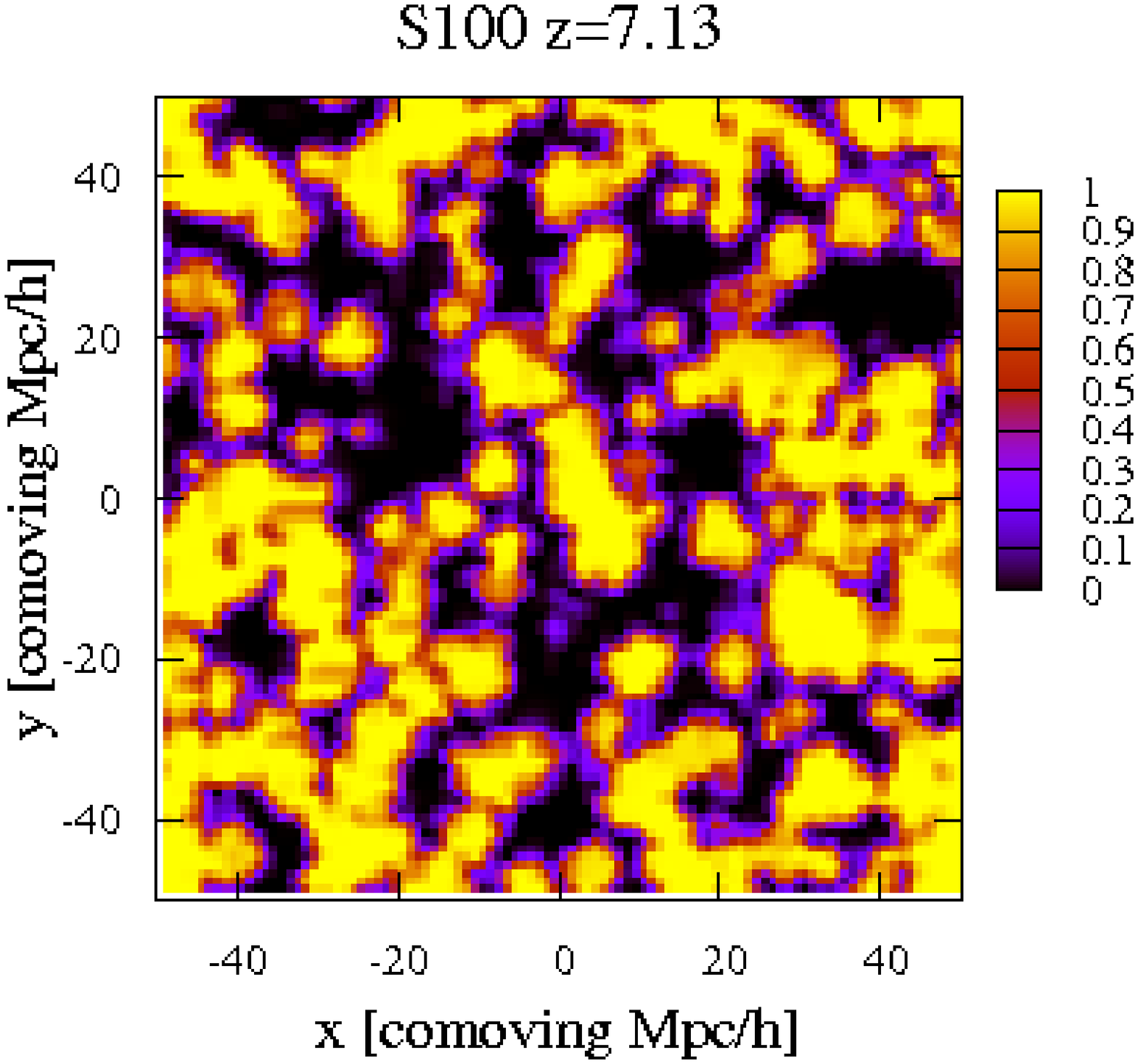}} 
\end{array}
\caption{Maps of the ionization fraction for the two simulations (right and left) at two different stages of reionization (above and below line), the WFCR and the HRR (see text for definition) and for two slice thicknesses (as labeled). The color scale is
a linear function of the ionization fraction.}
\label{ion_maps}
\end{figure*}

Maps of the ionization fraction of hydrogen are presented in Fig.~\ref{ion_maps} for both box sizes at a redshift when
$\langle x_\alpha\rangle_{\rm{vol}}=1$, which we will call the Wouthuysen-Field coupling redshift (hereafter WFCR), and at a redshift when 
$\langle x_H\rangle_{\rm{mass}}  =0.5$, which we will call the half reionization redshift (HRR). We can see that the WFCR correspond to
an early stage in the reionization history. The maps are presented 
for two different 
slice thicknesses~: $2\,h^{-1}\,\rm{Mpc}$ and $200\,h^{-1}\,\rm{kpc}$. The $2\,h^{-1}\,\rm{Mpc}$ slice corresponds to a bandwidth $\delta \nu \sim 0.15$ 
MHz at a redshift between $6$ and $10$, an acceptable value for SKA. The $200$ h$^{-1}$ kpc slice is plotted to give a 
feeling of how much power would be added at small scales by using narrower bandwidth~: indeed,  sharper 
ionization fronts are found in this case. Also as expected the S100 maps show
coherent structures on scales up to $50\,h^{-1}\,\rm{Mpc}$ which cannot be accounted for in the S20 maps. 

 Usual effect of a lower mass resolution (S100 vs S20) is to delay reionization. Indeed small mass halos form first and make a large
contribution to the ionizing photon budget at early times (see Iliev et al. \cite{Iliev07}). In our treatment, the delay is removed because 
we calibrated the source formation history of the S100 simulation on the S20 simulation by boosting the formation efficiency. However, this
procedure can only account for unresolved halos located within more massive, resolved, proto-halos. The contribution from unresolved halos
located in comparatively underdense regions could modify the morphology of the ionization maps during the early reionization history.

\begin{figure*}[t]
\centering
\resizebox{\hsize}{!}{\includegraphics[angle=270]{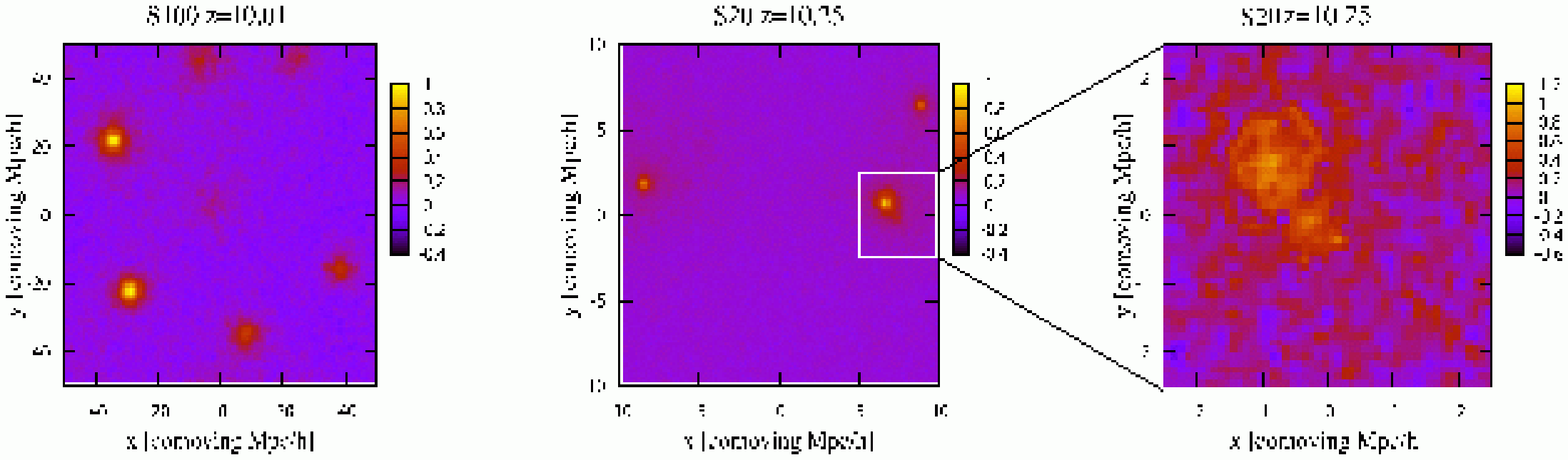}}
\caption{ Maps of the $x_\alpha$ coupling coefficient. We use a logarithmic color scale, and choose the redshift in each
simulation (S20 and S100) such that $\langle x_\alpha\rangle=1$ (10.25 for S20 and 10.01 for S100). The thickness of the slice is $2\,h^{-1}\,\rm{Mpc}$  for the
two panels on the left and $100$ h$^{-1}$.kpc for the right panel. }
\label{xalpha_maps}
\end{figure*}

\subsubsection{Lyman-$\alpha$ coupling}

The influence of including a proper treatment of the Wouthuysen-Field effect on the $21\,\rm{cm}$ signal is twofold. First, it 
simply modifies the intensity of the signal, especially in the regions seen in absorption. This will be obvious
in the brightness temperature maps shown in section 4.2.3. The second effect is to introduce a new source of fluctuations
in the signal, associated with the fluctuations in the local Ly-$\alpha$ flux. Fig.~\ref{xalpha_maps} presents maps of
the $x_\alpha$ coefficient for both simulations at the WFCR. 
Around the WFCR, the additional 
fluctuations to the brightness temperature are expected to be maximal. On a large scale, the maps boil down to strong coupling regions 
around sources, superimposed on a uniform weak coupling background. The signature of the added brightness temperature fluctuations will
be based on the typical sizes of the strong coupling regions and on the clustering and Poisson noise in the source
distribution. Keeping this in mind, it is interesting to notice that the strong coupling regions appear to be much smaller, $\sim 2\,h^{-1}\,\rm{Mpc}$, 
in the S20 simulation than in the S100 simulation where they extend over $\sim 10\,h^{-1}\,\rm{Mpc}$. To understand this
let us remember that we apply periodic boundary conditions to the radiative transfer code. As a result, any point in the
S20 simulation is surrounded by a rather homogeneous distribution of sources, with the closest source at most 
$\sim 10\,h^{-1}\,\rm{Mpc}$ away. This dampens the fluctuations in the Ly-$\alpha$ flux and washes out the outer parts of the
strong coupling regions.  As a result, we expect to find more power at large scales in the added brightness temperature fluctuations in the
S100 simulation than in the S20 simulation. 

The zoom on the source in the S20 simulation shows that the apparent spherical symmetry of the strong coupling regions 
breaks down at small scales. Semelin et al. (\cite{Semelin07}) have shown that substantial asymmetry can be expected if
the density field of the gas is not homogeneous (e.g. presence of filaments). Let us notice, however, that the slice
thickness for the zoom is $100$ h$^{-1}$ kpc, while it is $2\,h^{-1}\,\rm{Mpc}$  for the other maps. With such a thin slice
the noise in the $x_\alpha$ evaluation (resulting of the finite number of photons used in the Monte Carlo scheme) 
is not smoothed out at all. Any structure outside the main strong coupling region is mostly noise (visible on the color version only). 
If structures exist in the
$x_\alpha$ maps at scales smaller than $1\,h^{-1}\,\rm{Mpc}$  and with an amplitude of less than a factor of $3$ above or below
the background, we will miss them in the noise or smooth them out. From a theoretical point of view it would be interesting
(but very costly) to produce maps with a lower noise level. However, from a practical point a view, these scales are below the
projected resolution of SKA.

\subsubsection{Brightness temperature maps}

\begin{figure*}[p]
\centering
\begin{array}[t]{|c|c|c|}
\hline
\vspace{-0.1cm}
\hbox{\bf \large Case 1~: $ T_S \gg T_{CMB} $} &
\hbox{\bf \large Case 2~: $ T_S = T_{K} $} &
\hbox{\bf \large Case 3~: $ T_S(x_\alpha) $} \\
\hline
\vspace{-0.2cm}
\resizebox{5.4cm}{!}{\includegraphics[angle=270]{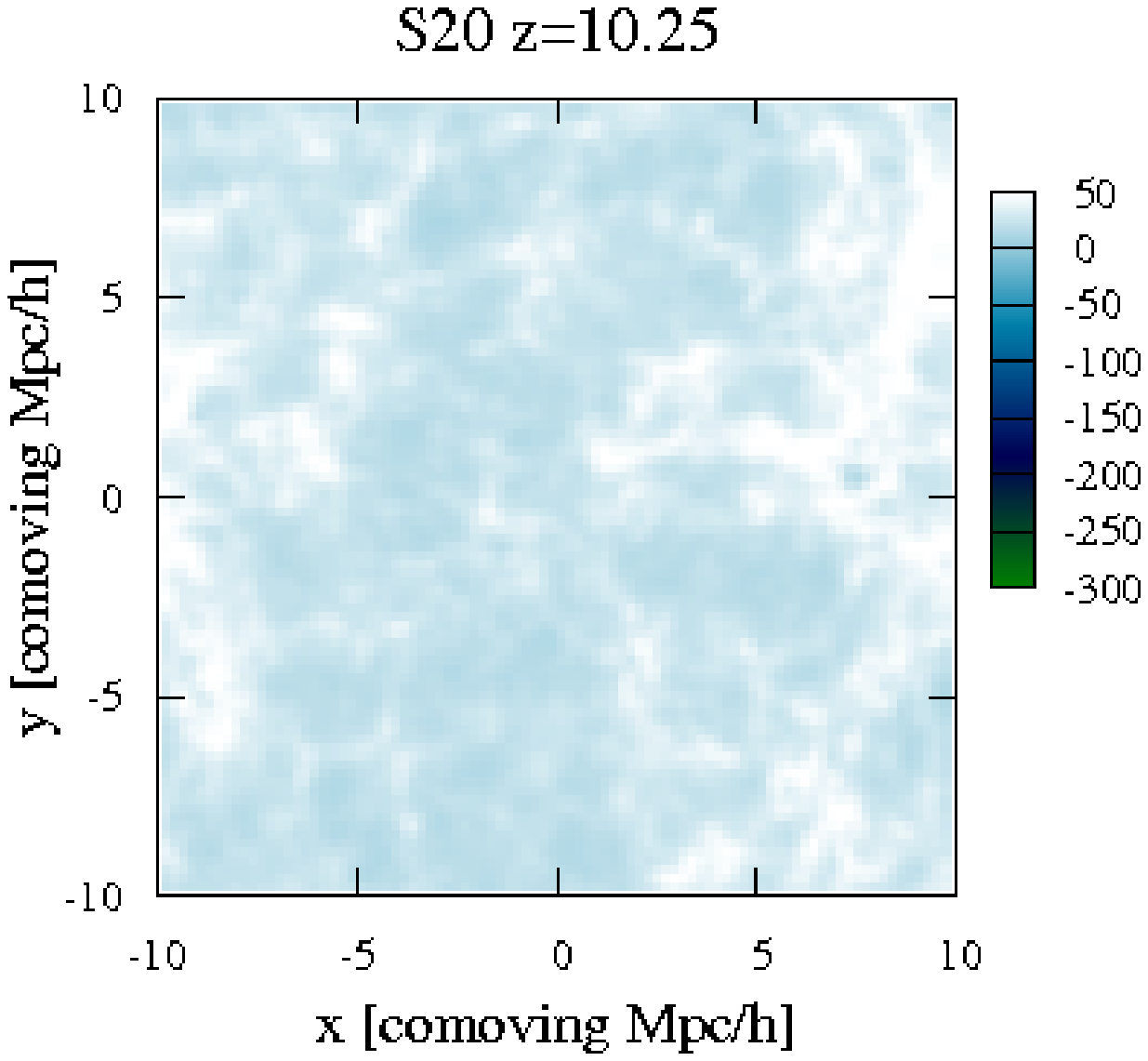}} \hskip 0.3cm &
\hskip 0.3cm \resizebox{5.4cm}{!}{\includegraphics[angle=270]{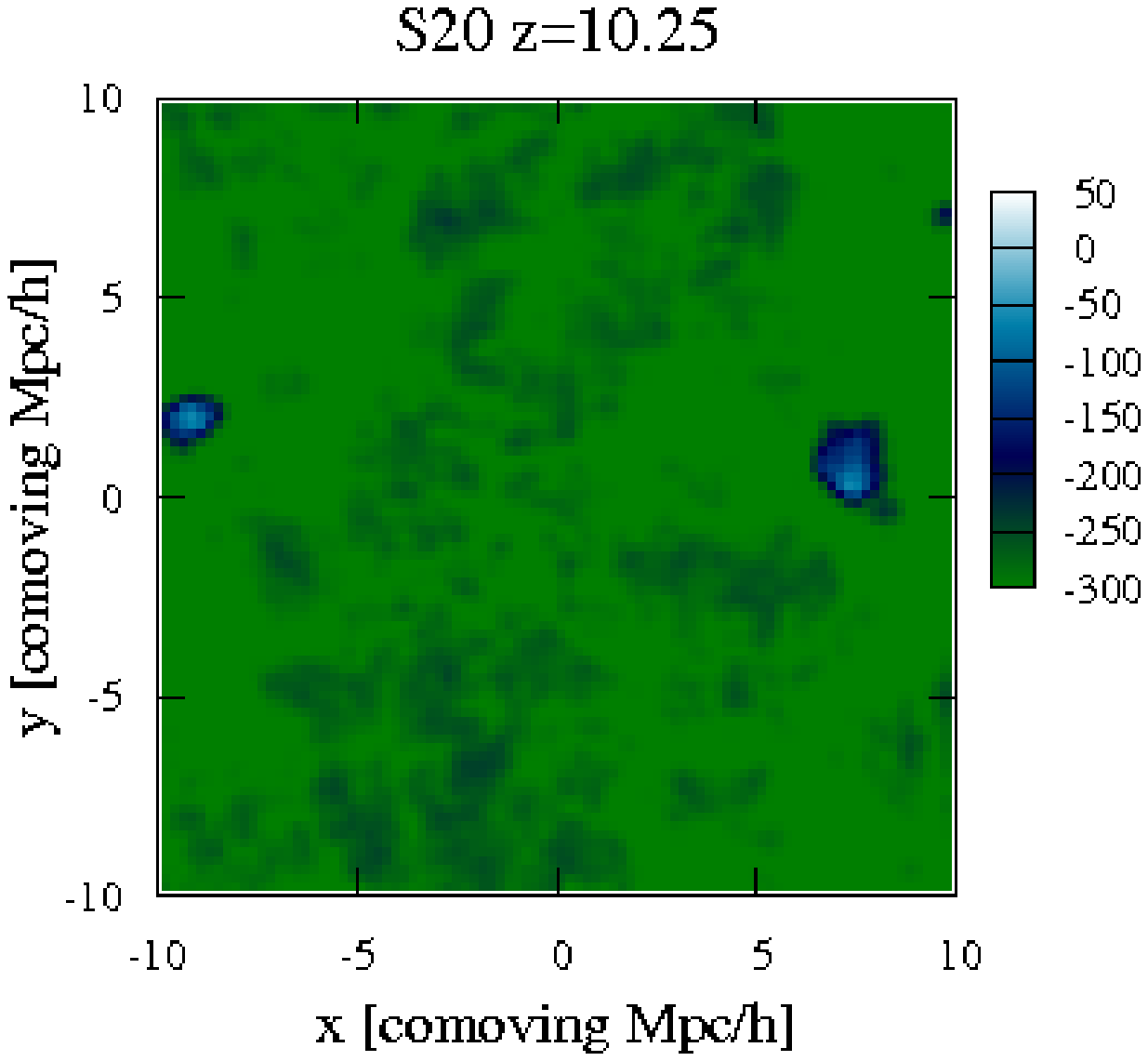}}  \hskip 0.3cm&
\hskip 0.3cm \resizebox{5.4cm}{!}{\includegraphics[angle=270]{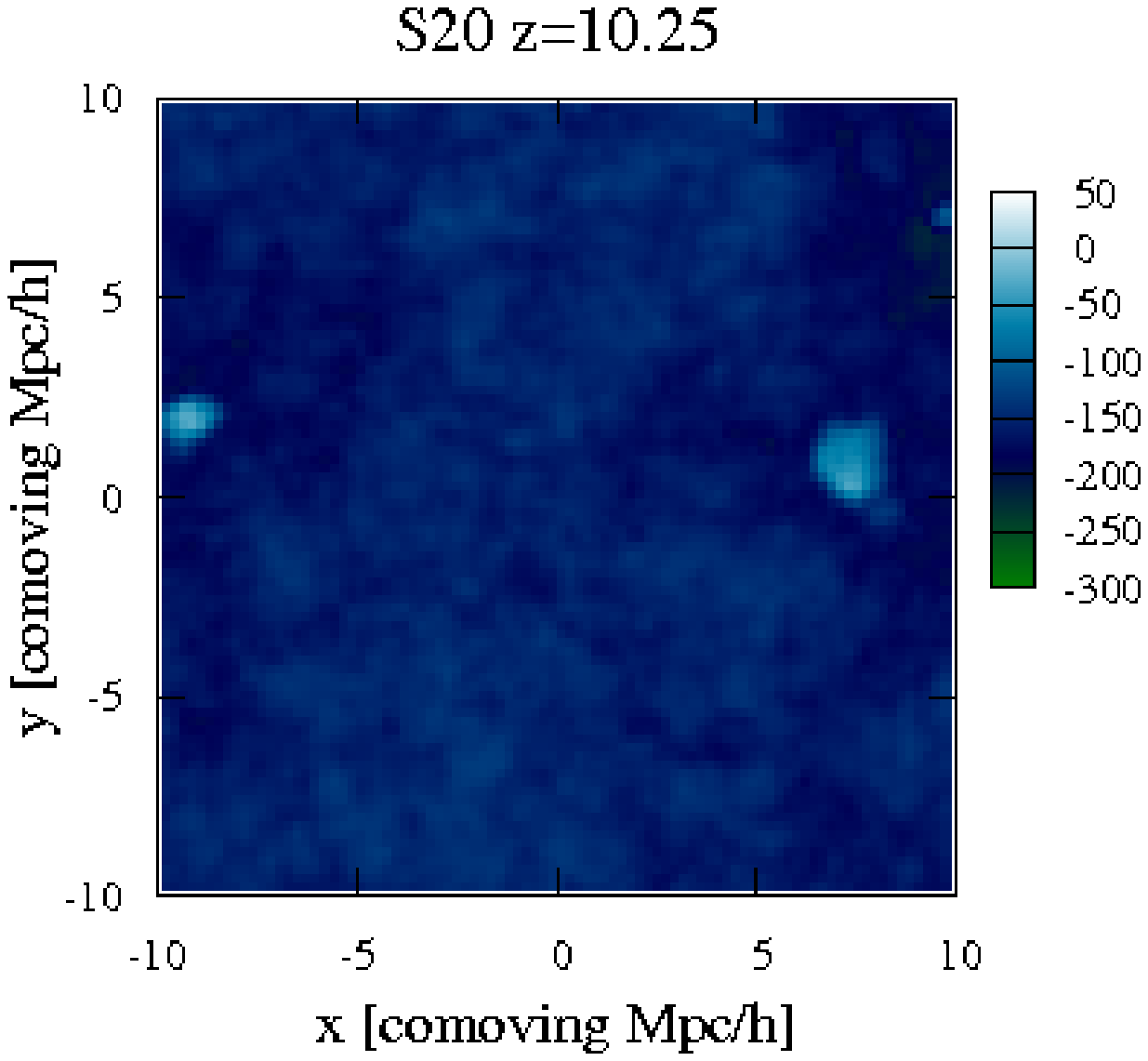}} \\

\vspace{0.1cm}
\resizebox{5.4cm}{!}{\includegraphics[angle=270]{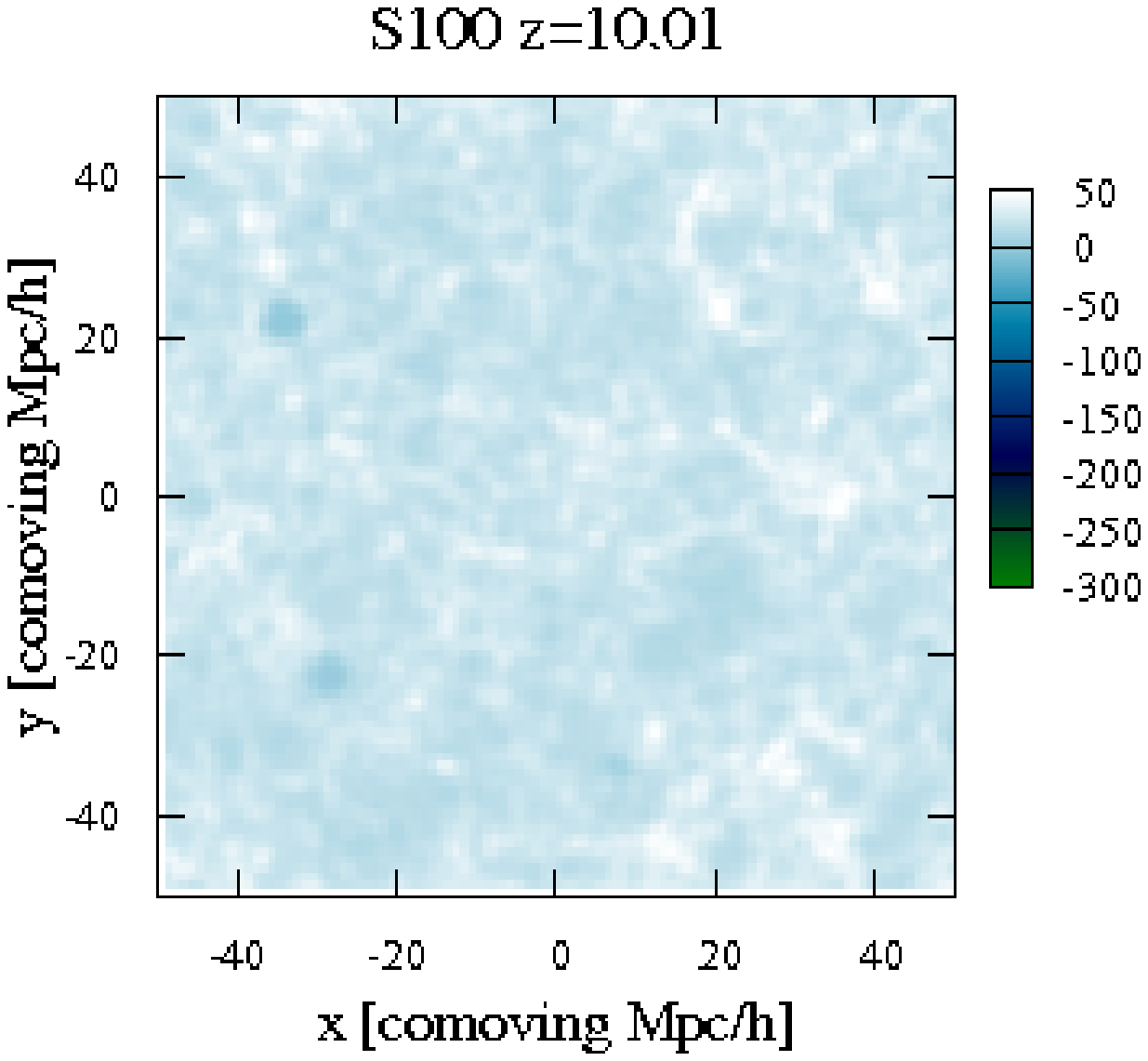}} \hskip 0.3cm& 
\hskip 0.3cm \resizebox{5.4cm}{!}{\includegraphics[angle=270]{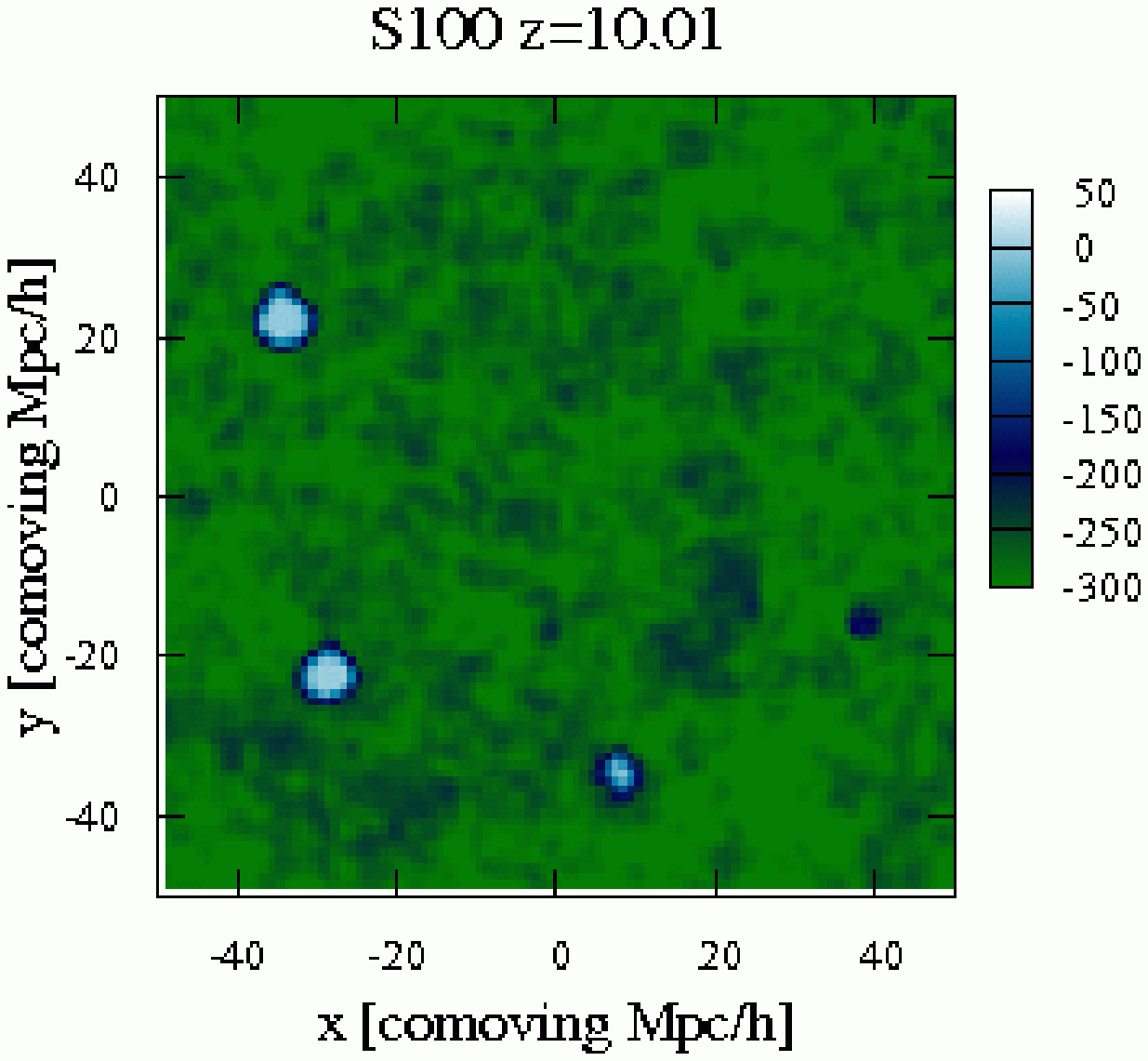}} \hskip 0.3cm&
\hskip 0.3cm \resizebox{5.4cm}{!}{\includegraphics[angle=270]{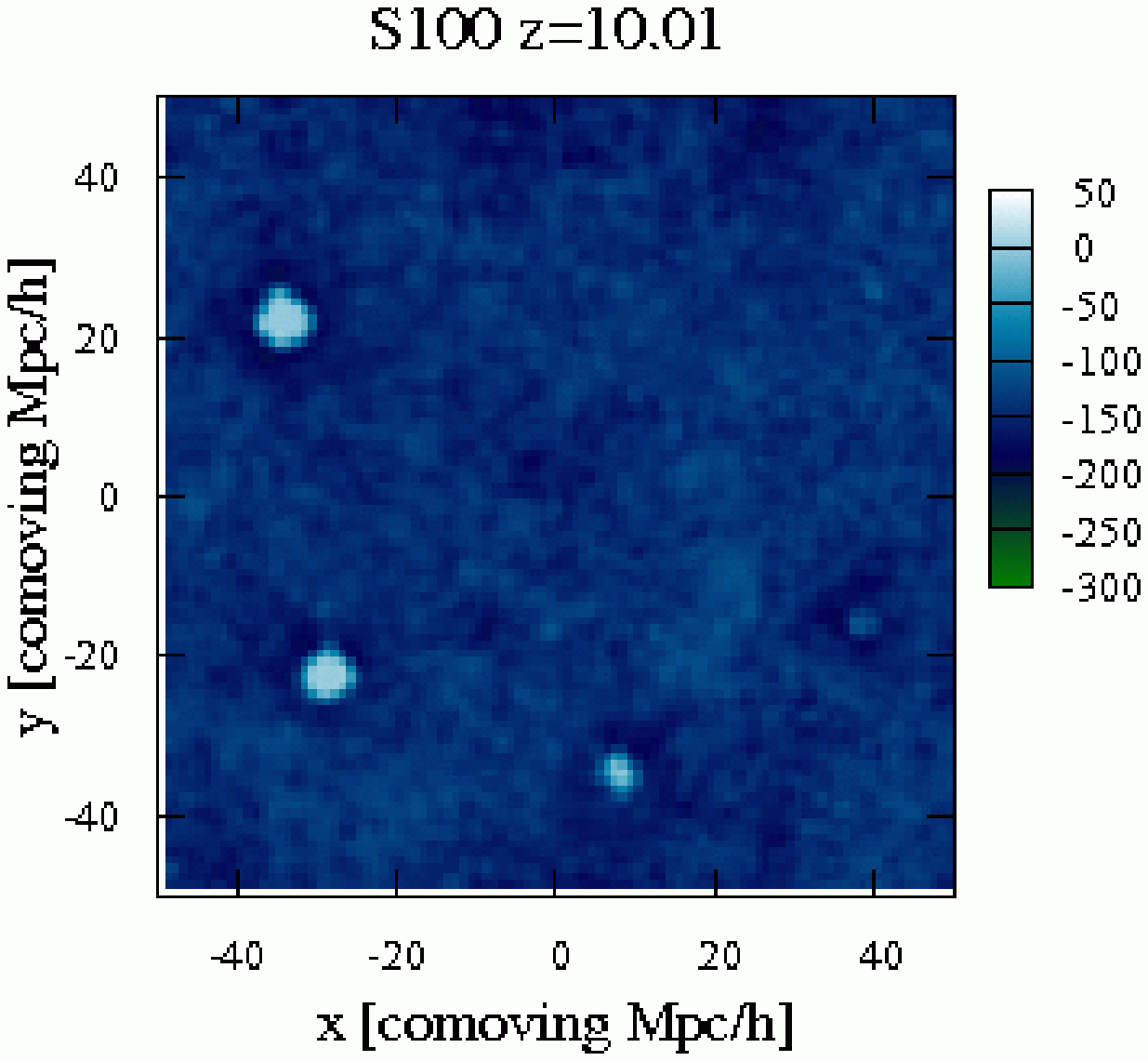}} \\
\hline
\vspace{-0.2cm}
\resizebox{5.4cm}{!}{\includegraphics[angle=270]{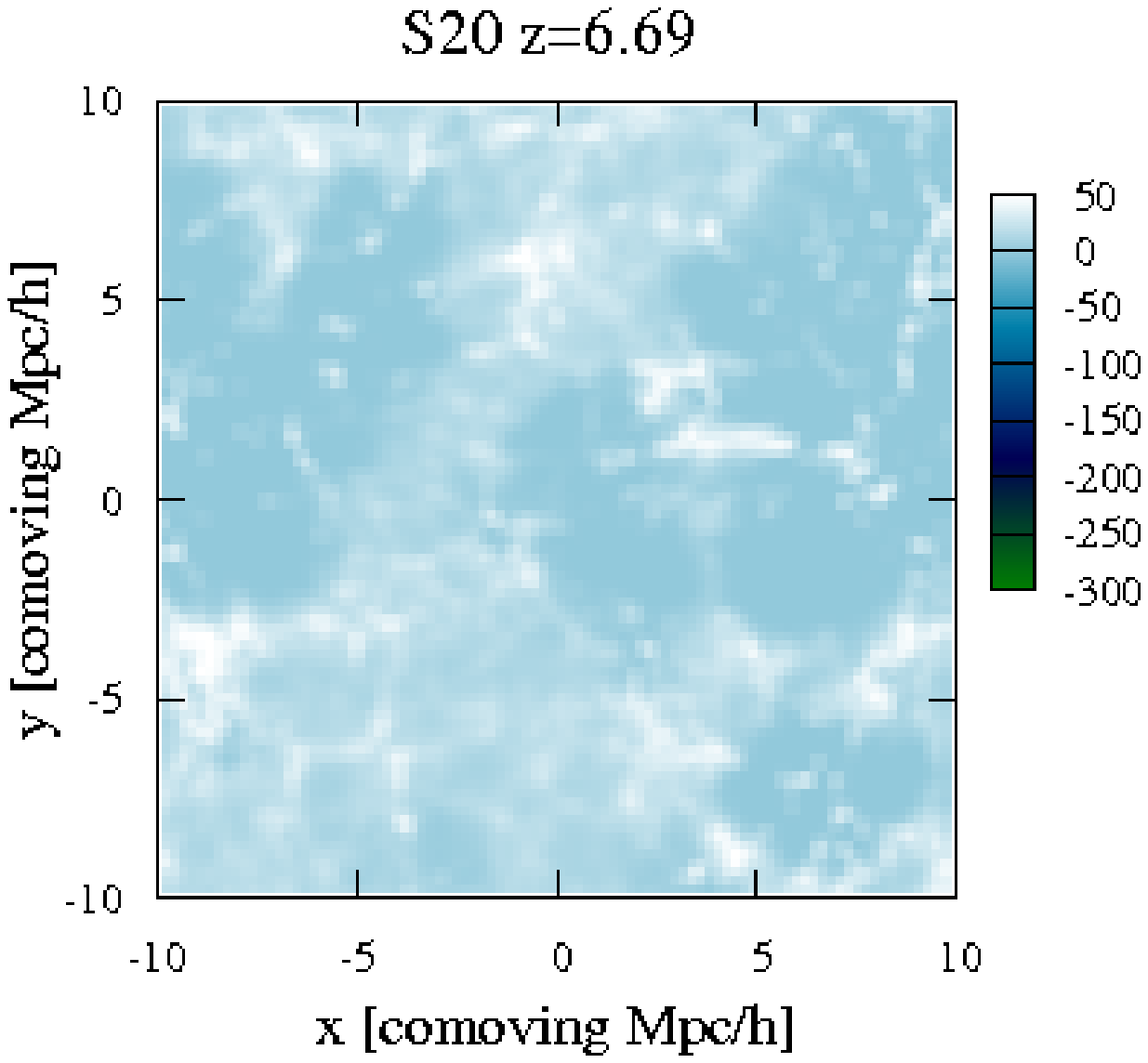}} \hskip 0.3cm &
\hskip 0.3cm \resizebox{5.4cm}{!}{\includegraphics[angle=270]{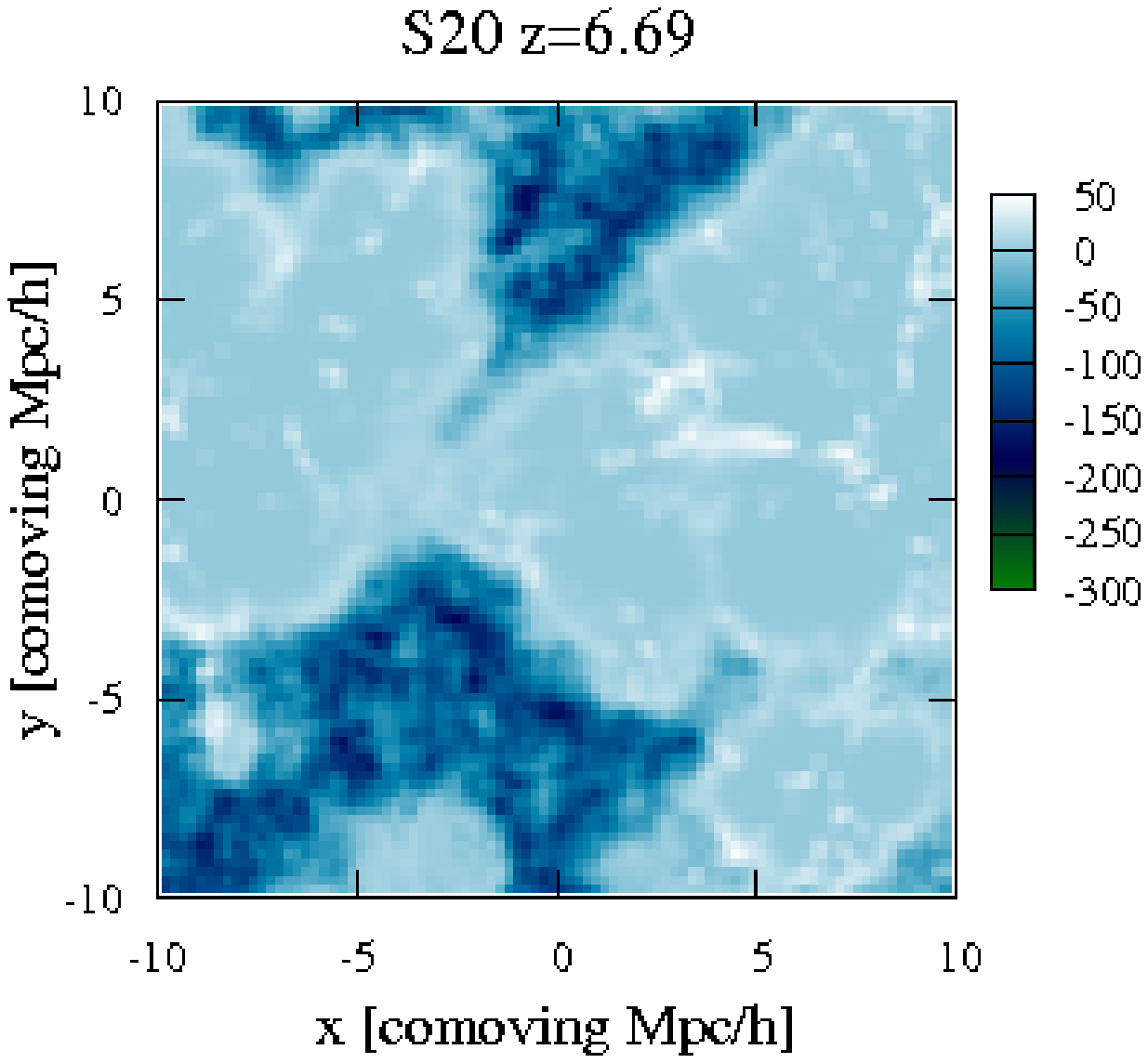}} \hskip 0.3cm &
\hskip 0.3cm \resizebox{5.4cm}{!}{\includegraphics[angle=270]{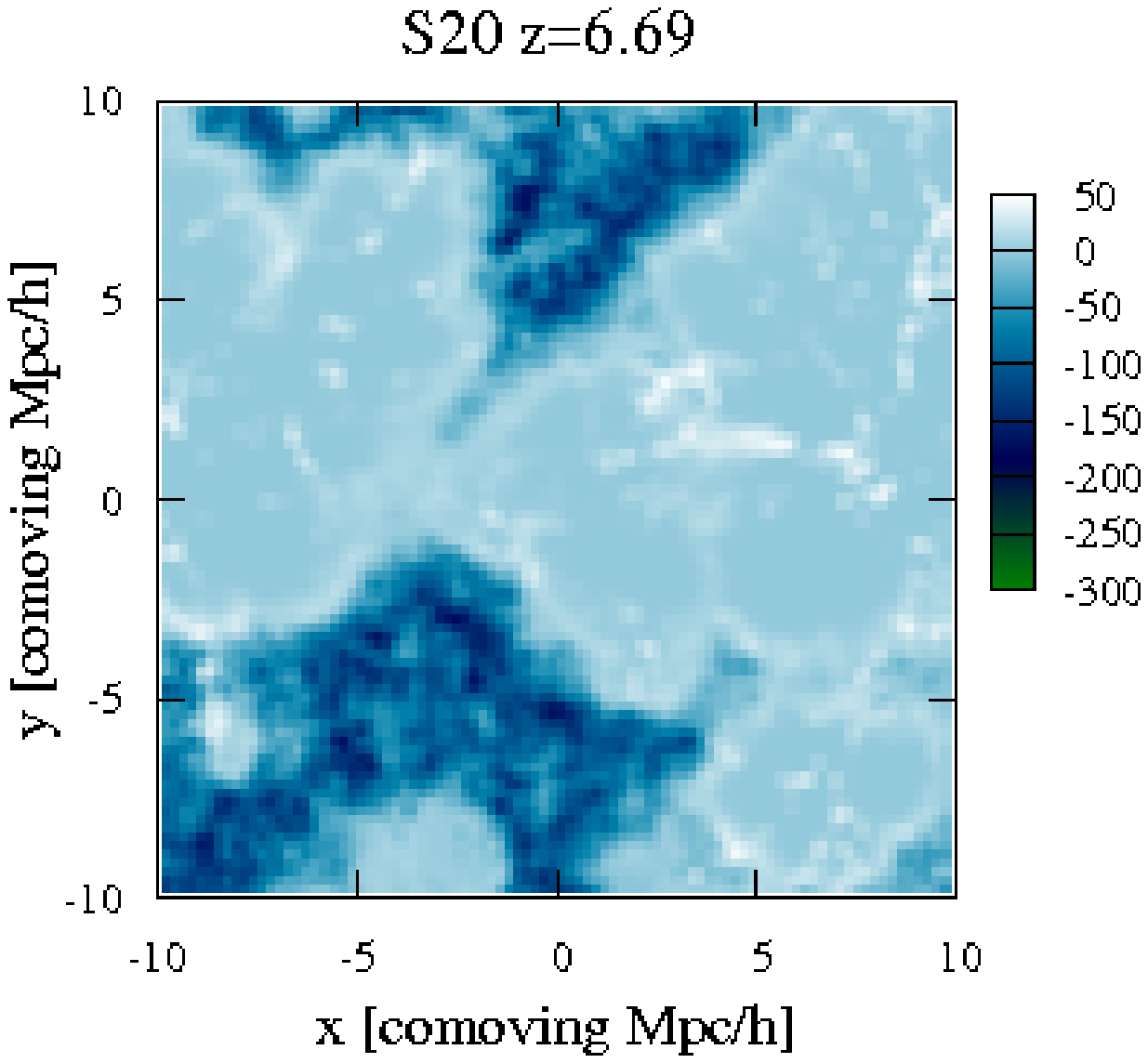}}\\

\resizebox{5.4cm}{!}{\includegraphics[angle=270]{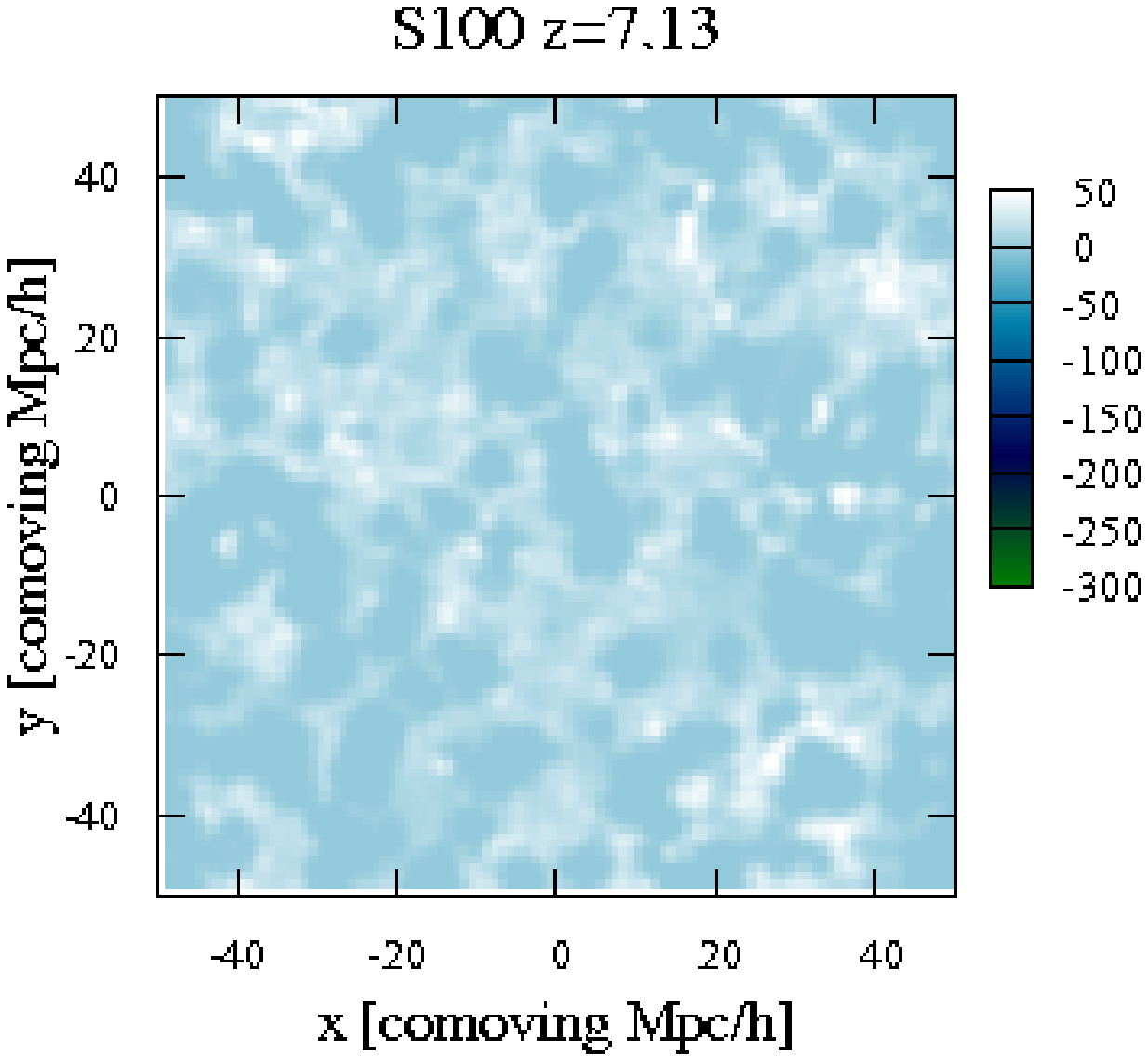}} \hskip 0.3cm &
\hskip 0.3cm \resizebox{5.4cm}{!}{\includegraphics[angle=270]{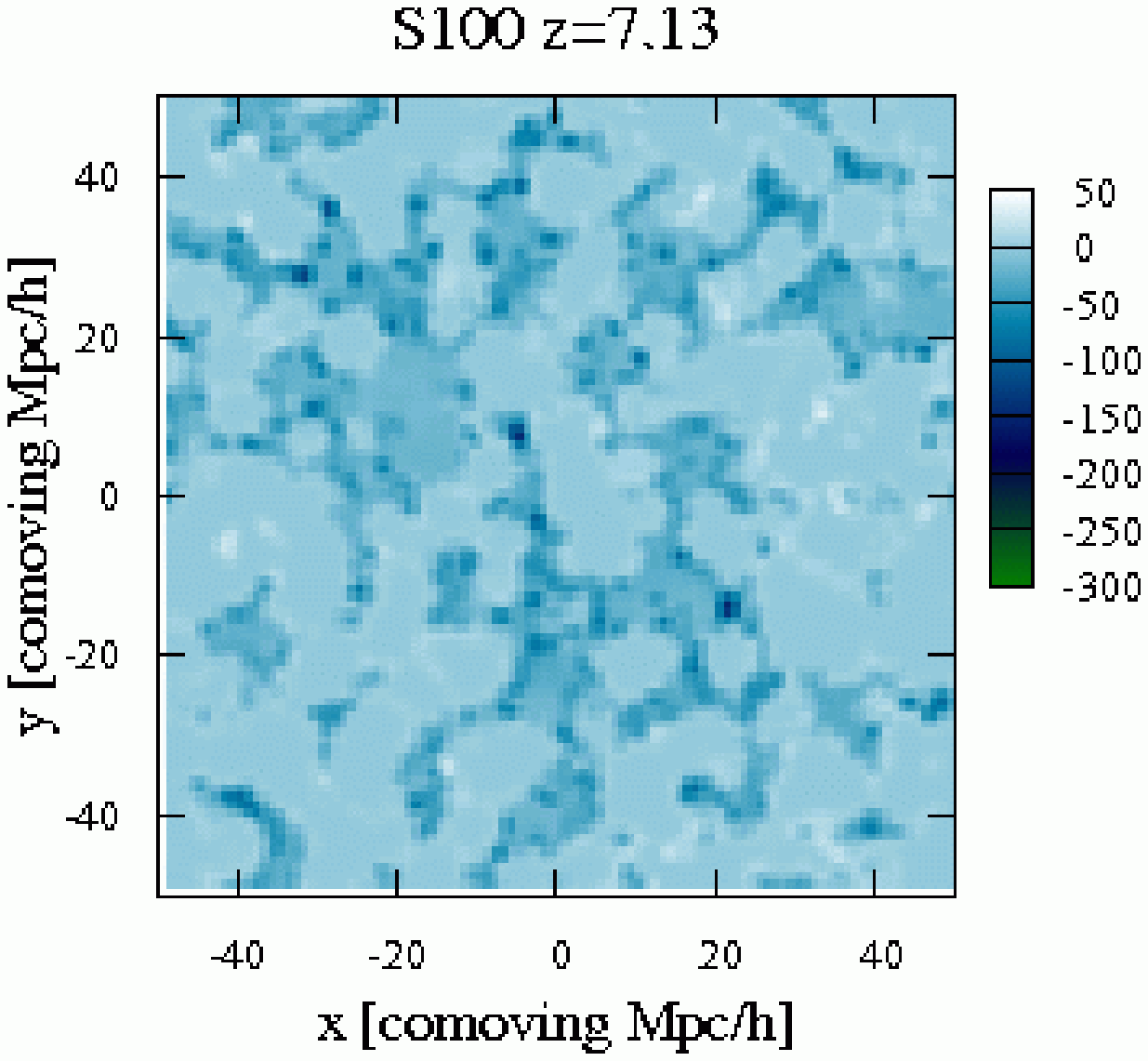}} \hskip 0.3cm &
\hskip 0.3cm \resizebox{5.4cm}{!}{\includegraphics[angle=270]{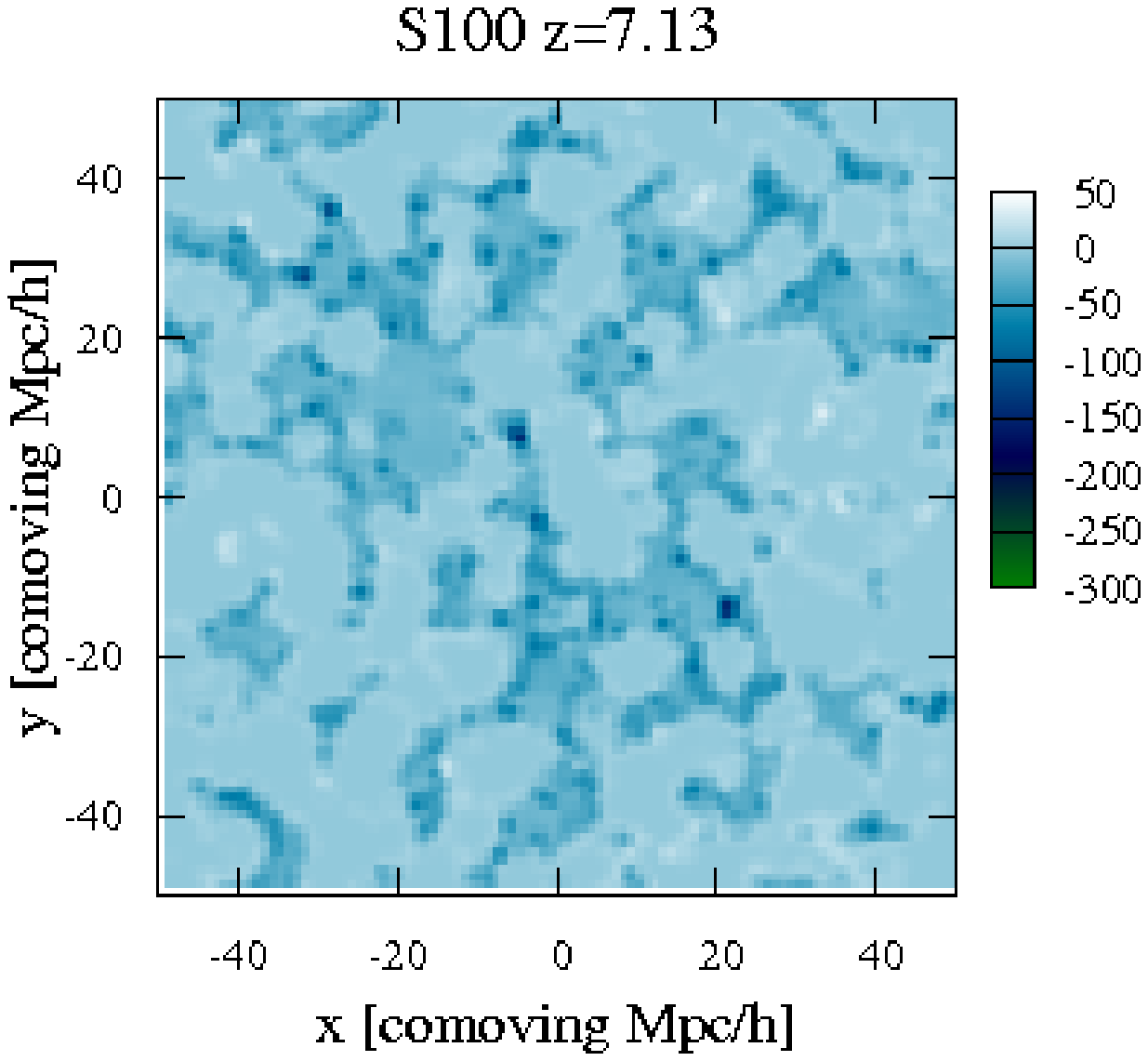}} \\
\hline

\end{array}
\caption{Differential brightness temperature maps for both box sizes and 3 different hypotheses for computing the spin 
temperature (see main text and description in the figure). The redshift of the 6 top panels is chosen such that 
$\langle x_\alpha \rangle =1$, the redshift of the 6 bottom panels is such that $\langle x_H \rangle =0.5$. The temperature color scale is linear, in mK. The thickness of the slice is $2\,h^{-1}\,\rm{Mpc}$ . }
\label{deltaTb_maps}
\end{figure*}

The most common (and most drastic) approximation used to compute the brightness temperature is $T_S \gg T_{CMB}$ (case 1).
 In this
approximation no absorption region can exist. The second approach is to assume full Lyman-$\alpha$ coupling ($x_\alpha =
\infty$) (case 2). The third approach, the one used in this paper, is to compute self-consistently the local values of 
$x_\alpha$ and derive accordingly the local values of $T_S$ and $T_b$ (case 3). We plot maps of the differential 
brightness temperature for all three cases and for both simulation boxes at the WFCR in the 6 top panels of Fig.~\ref{deltaTb_maps}. 
The thickness of the slice is $2\,h^{-1}\,\rm{Mpc}$ . 
At this early redshift, when the two first approximations are hardly reliable, the differences are striking. Not
surprisingly case 1 fails completely~: it is not tailored to handle absorption regions and predicts only emission. More
interesting, but also expected, case 2 overpredicts the strength of the absorption compared to case 3 ($300\,\rm{mK}$ instead of $200\,\rm{mK}$) by assuming full Wouthuysen-Field coupling. We used the same color scale for all maps to emphasize these
differences. The strong absorption signal observed in cases 2 and 3 is the consequence of our choice
of rather soft-spectrum sources. The neutral IGM undergoes very little pre-heating by UV and X rays. Let us emphasize
that the strength of the absorption signal is very sensitive to the thermal modeling of the gas when $T_K
\ll T_{CMB}$, which is not uncommon at this redshift with our choice of sources. Indeed, the included Ly-$\alpha$ 
heating, although very weak (at most a few K over the EoR), decreases the strength of the absorption signal by $50$ to 
$100\,\rm{mK}$. If even a small fraction of hard-spectrum sources was included, the absorption strength would decrease and turn to emission at later redshifts .
 
The six bottom panels of Fig.~\ref{deltaTb_maps} shows the same maps for the HRR. At this redshift we have $ \langle 
x_\alpha \rangle > 200 $ for both simulations. Consequently the maps for cases 2 and 3 are almost identical. Comparing
with Fig.~\ref{ion_maps} it can be checked that the regions which are still seen in absorption are the neutral regions not too
close to the ionization fronts. Comparing the absorption regions in the S20 and in the S100 maps, it can be seen that the signal
is somewhat stronger in the S20 simulation. We have determined that a lower gas kinetic temperature in the S20 simulation, in the voids,
is responsible for this effect. Since we are talking about temperatures lower than $10$ K, a difference of a few K
produce a large effect on the differential brightness temperature. The S20 simulation has a higher mass resolution which creates higher
density contrasts. In the voids, the density is lower and the adiabatic expansion stronger~: the resulting temperature is lower.

Obviously, in case 1, the absorption regions turn to emission. Since emission saturates quickly at a few tens of mK, the
statistical property of the signal such as the average rms fluctuations of the brightness temperature will be weaker. At such
an advanced stage in the reionization history the assumption made in case 1 may actually be more realistic: case 2 and 3 would
probably look more like case 1 if we included X-ray sources.

\begin{figure}[t]
\centering
\resizebox{\hsize}{!}{\includegraphics{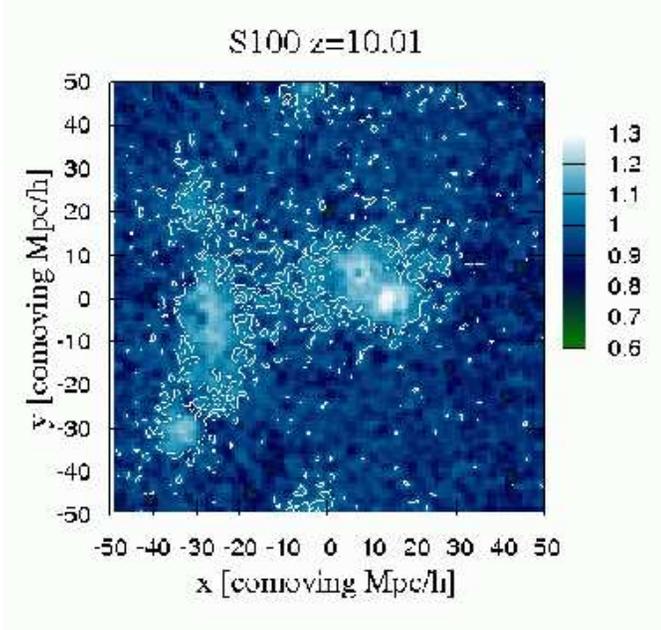}}
\caption{ Ratio of the differential brightness temperature computed with the fluctuating local
value of $x_\alpha$ on the one hand and the average value of $x_\alpha$ on the other hand. The color scale is linear. The contour is computed 
from the $x_\alpha$ field for the $x_\alpha=1$ value. }
\label{dTbratio}
\end{figure}

Using a homogeneous average value for $x_\alpha$ to account for the Wouthuysen-field effect ( for example a calibrated function of $x_H$)
 is much easier than computing the full radiative transfer in the line. So it is important to assess the benefit reaped from doing the full treatment. The $x_\alpha$ field induces its
own fluctuation in the brightness temperature signal. Fig.~\ref{dTbratio} illustrates this point. It shows  a map of the ratio of the 
differential brightness temperatures computed with the fluctuating local value of $x_\alpha$ on the one hand and the homogeneous, average value 
of $x_\alpha$ on the other hand. This map is computed at the WFCR, for the S100 simulation. As expected, 
 the ratio is mostly larger than one inside the $x_\alpha=1$ contour and lower outside. Moreover the two 
differential brightness temperatures differ by up to 30\%. Consequently it appears that, at these early redshifts when the Lyman-$\alpha$ 
coupling is not full, it is worth doing the full Lyman-$\alpha$ transfer.

\begin{figure}[t]
\centering
\includegraphics[scale=0.3,angle=-90]{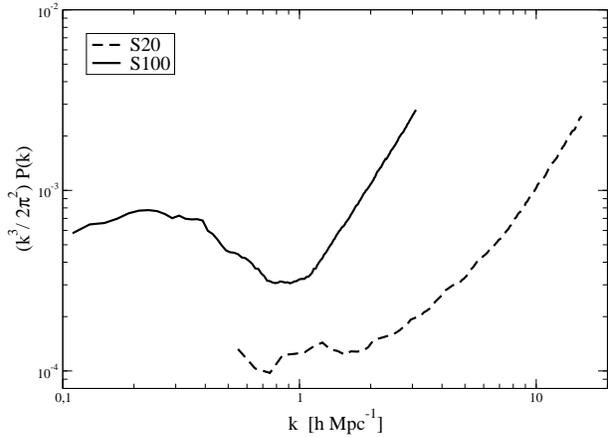} \\
\caption{3D isotropic power spectrum of ${x_\alpha \over 1+x_\alpha}$. The spectra are plotted for both simulations, S20 and S100, at the redshift when $\langle x_\alpha \rangle =1$.}
\label{x_alpha_powerspec_2boxsizes}
\end{figure}

Using the power spectrum as a statistical tool for analyzing the EoR $21\,\rm{cm}$ signal is natural because it can be directly 
computed from the visibilities measured by the radio interferometers such as LOFAR, MWA or SKA. 
Morales \& Hewitt (\cite{Morales04}) have shown that using the full 3D power spectrum rather than the 2D angular 
power spectrum will improve the sensitivity of the observations and make the process of foreground removal easier.

\subsubsection{Strength of the Lyman-$\alpha$ coupling fluctuations}

Fig.~\ref{x_alpha_powerspec_2boxsizes}
shows the power spectrum of the quantity  $z_\alpha={x_\alpha \over 1+x_\alpha}$ for both simulations at the WFCR. Indeed directly plotting the spectrum of $x_\alpha$ is not interesting~: the $x_\alpha$ field is dominated by the very 
strong values near the sources and has a power spectrum similar to that  of a distribution of Dirac functions. However, these
very high values of $x_\alpha$ are not especially interesting~: they are just values for which the coupling saturates. 
$z_\alpha$, which is just the weight applied to the inverse kinetic temperature in Eq.~$9$ (neglecting the contribution of $x_c$),
varies between $0$ and $1$ and contains more relevant information about the magnitude of the coupling. As  can be seen
in Fig.~\ref{x_alpha_powerspec_2boxsizes}, the point sources still dominate the small scales. More interesting is the
increase of the power spectrum toward large scales, which is barely visible for the S20 simulation but obvious in the S100
simulation. It is likely that this increase tracks the Poisson noise in the large scale clustering of the sources 
and void distribution.
For the S100 simulation, the spectrum shows a local maximum at scales $\sim 30\,h^{-1}\,\rm{Mpc}$. Whether this maximum is 
real or only the effect of the box size is an interesting question. Only simulations with larger boxes will give the
answer.

\begin{figure}[t]
\centering
\begin{array}[t]{c}
\includegraphics[scale=0.3,angle=-90]{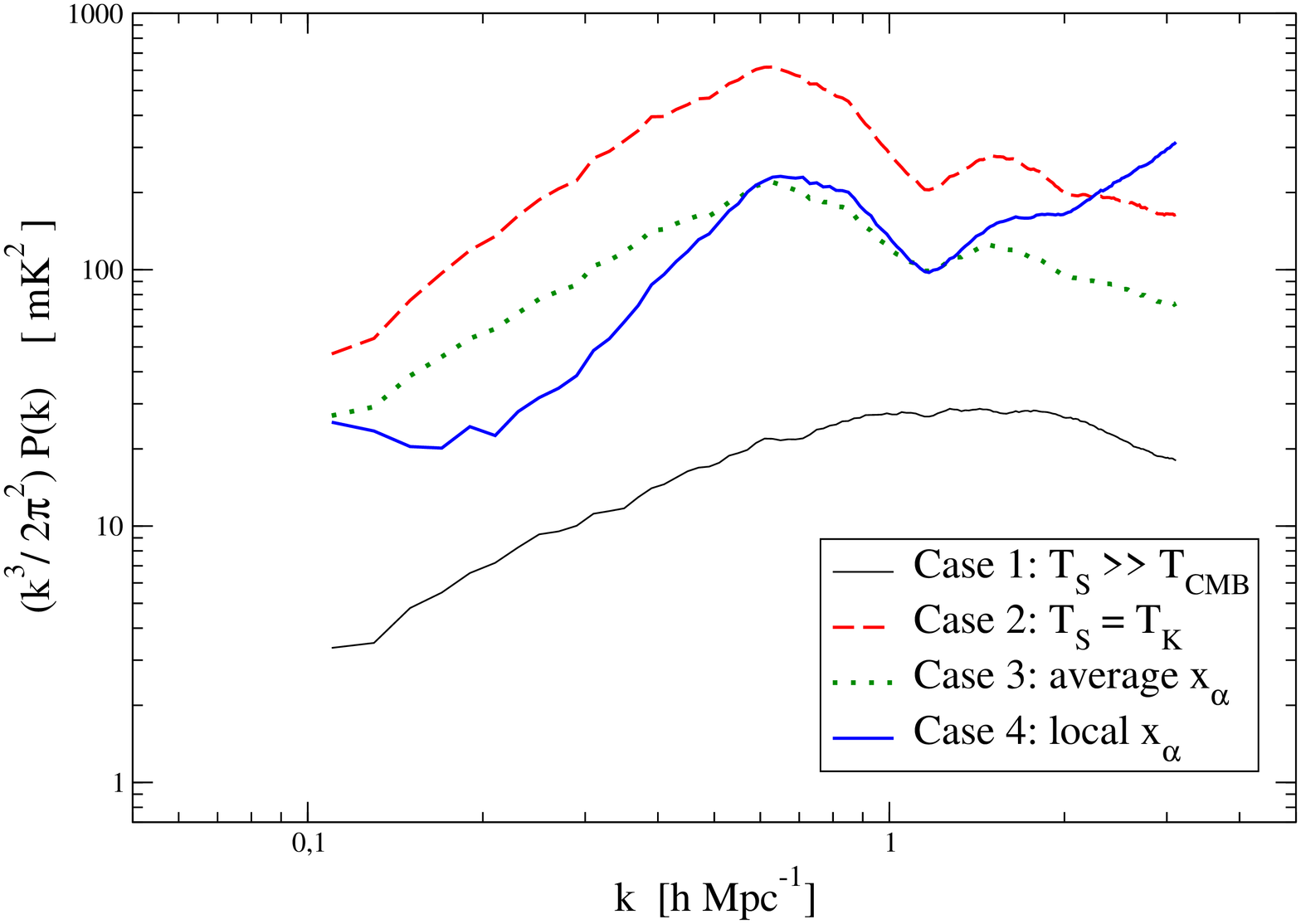} \\
\includegraphics[scale=0.3,angle=-90]{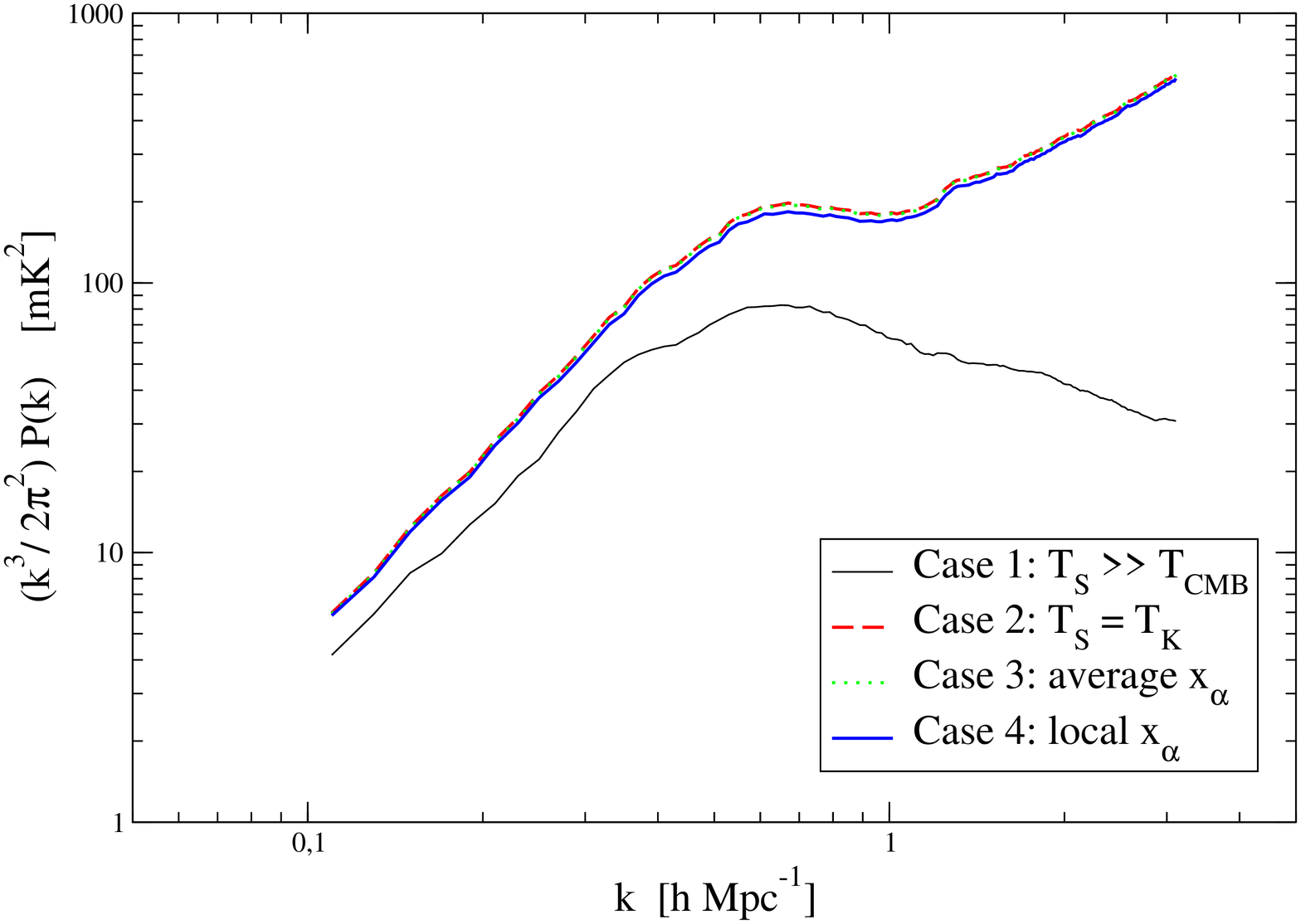} \\
\end{array}
\caption{Comparison of the 3D isotropic dimensional power spectrum of the differential brightness temperature with 4 
different assumptions used to compute the spin temperature (see legend and main text). The spectra are computed for the
S100 simulation. The top panel is for the redshift when $\langle x_\alpha \rangle =1.$ and the bottom panel is for 
the redshift when $\langle x_H \rangle =0.5$.}
\label{dTbpowspec_4cases}
\end{figure}

\subsubsection{Differential brightness temperature}
In this section, we will compute directly the {\sl dimensional} power spectrum of $\delta T_b$.
Various authors, studying the emission regime, compute the power spectrum of  $\delta T_b$ divided by the 
average, redshift dependent, emission temperature 
$T_0 \sim 28 \left({1+z \over 10}\right)^{1/2}$. When
we include the absorption regime, $T_0$ has no special significance. Using the actual average $\delta T_b$ 
of the box would be no better~: having both emission and absorption, it holds no information about the 
amplitude of the fluctuations and may even be zero at some redshift. Using the actual rms average at each redshift would 
indeed be more relevant as it does reflect the amplitude of the fluctuations. However it would erase the information
about the absolute amplitude at a given scale, at different redshifts. Consequently, we decided to plot the direct 
dimensional power spectrum of the differential brightness temperature.

\begin{figure*}[t]
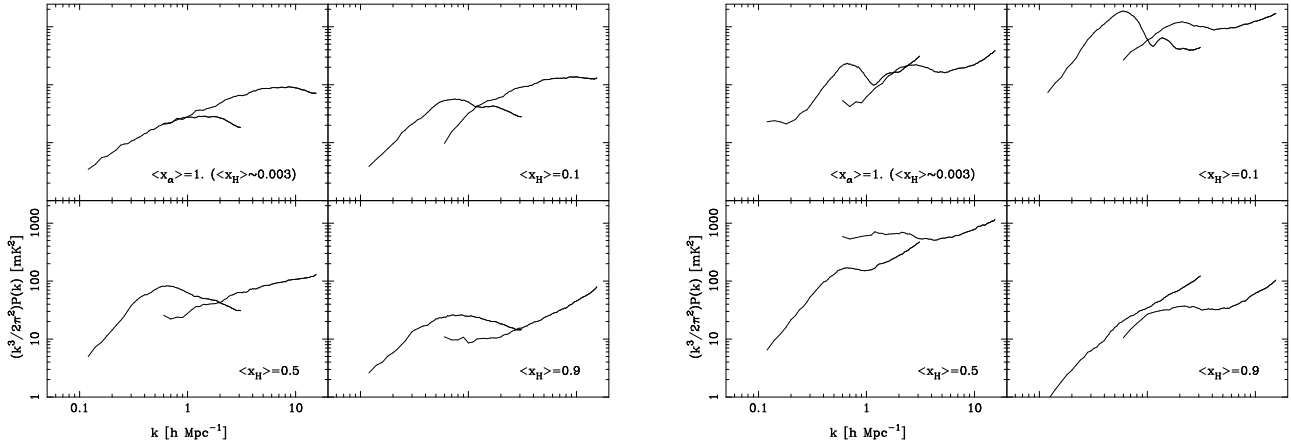

\centering
\renewcommand{\arraycolsep}{0mm}
\begin{array}[t]{cc}
\includegraphics[scale=0.7,angle=-90]{powerspec/left_panel.ps}  \hskip 1cm &
\includegraphics[scale=0.7,angle=-90]{powerspec/right_panel.ps}

\end{array}
\caption{3D isotropic dimensional power spectrum of the differential brightness temperature for the two box sizes. The
stage of reionization is indicated in each panel. The four left panels assume $T_S \gg T_{CMB}$ while the four right
panels compute $T_S$ exactly using the local values of $x_\alpha$.}
\label{dTbpowspec_2boxes}
\end{figure*}

Fig.~\ref{dTbpowspec_4cases} shows the three-dimensional isotropic power spectrum of $\delta T_b$ as 
defined by Eq.~$8$ for the S100 simulation, at the WFCR (top panel) and the HRR (bottom panel), 
in four different cases for the computation of
the spin temperature. In case 1, we assume that $T_S \gg T_{CMB}$, in case 2 $T_S=T_K$, in case 3 we use Eq.~$9$ with the box
averaged value of $x_\alpha$, and finally in case 4,  we use Eq.~$9$ with the local values of $x_\alpha$. As expected, the larger
differences appear at the WFCR, where the amplitude of the power spectrum changes by more than a factor of 10 depending 
on the case. Case 1, ignoring the possibility of absorption, strongly underestimates the amplitude of the power spectrum, 
while case 2, assuming full coupling between spin temperature and gas kinetic temperature, overestimates the amplitude.
The amplitudes for cases 3 and 4 are similar (and more realistic). However, case 4 shows more power at small scales and
less power at large scales than case 3. Indeed, as illustrated in Fig.~\ref{dTbratio}, case 3 uses a larger value of
$x_\alpha$ far from the sources, thus produces a stronger absorption in large scale voids and more power at large scale
in the spectrum than case 4. The behaviour at small scales arises from the presence of the Ly-$\alpha$ halos around
the sources in case 4. Semelin et al. (\cite{Semelin07}) have shown that the profile of these halos can be as steep as
$r^{-3}$ if the gas has an isothermal profile. Even at distances of a few comoving Mpc, where the gas profile flattens,
the Ly-$\alpha$ halo profile behaves as $\sim r^{-2.3}$ (Semelin et al. \cite{Semelin07}, Chuzhoy \& Zheng 
\cite{Chuzhoy07}). The contribution from these sharp $x_\alpha$ fluctuations explains the increase in the case 4 
power spectrum at small scales. At the HRR (bottom panel) the situation is simpler. The Wouthuysen-Field coupling
is very strong and case 2, 3 and 4 are almost indistinguishable. All three have much larger amplitude than case 1 
which is related to the fact that absorption can produce a larger absolute signal than emission, thus larger fluctuations.
It is interesting to notice that the difference in amplitude is especially large at small scales. The contribution of
the regions straddling the sharp ionization front is presumably responsible for this feature. In case 1, the brightness 
temperature rises from $\sim 0\,\rm{mK}$ in the fully ionized region inside the ionization front to a moderate $\sim 20\,\rm{mK}$
emission outside of the front. In all other cases, the temperature first rises to a moderate emission in the regions just
outside the ionization front, preheated above the CMB temperature by the hard part of the source spectrum, then drops to 
strong absorption farther out in the cool voids. The thickness of the preheated regions depends strongly on the source 
modeling~: in our case it is less than $1\,h^{-1}\,\rm{Mpc}$ . With a harder spectrum these regions would be thicker,  and
with X-ray sources they would be erased and the corresponding feature absent in the power spectrum.

Fig.~\ref{dTbpowspec_2boxes} shows the $\delta T_b$ power spectrum of both the S20 and S100 simulations at four different
stages of reionization (the WFCR,  $\langle x_H \rangle = 0.1$, $\langle x_H \rangle = 0.5$, and
$\langle x_H \rangle = 0.9$), for case 1 (the four panels on the left) and case 4 (the four panels on the right). Let us
first comment on how the spectra for the two simulations connect. In most cases the large scales in the S20 simulation have
less power than the same scales in the S100 simulation. This is the reason why EoR simulations have to use large boxes~: to
sample well enough the large scale clustering properties of the source distribution. There is however one exception,
the $\langle x_H \rangle = 0.5$ (HRR) in case 4, where the S20 simulation has more power even on its large scales. This
is related to the fact, already discussed in section 4.2.3, that the absorption in the voids is stronger in the 
S20 simulation and thus the fluctuations are amplified. A second point is that at the WFCR the case 4 spectra have a
more complex signature and a larger amplitude than the case 1 spectra which merely track the density fluctuations. Taking 
absorption  and Ly-$\alpha$ coupling correctly into account is essential at this early redshift. A third point is that
we reproduce the behaviour noted by Mellema et al. (\cite{Mellema06b}) and Lidz et al. (\cite{Lidz07b}), that at scales 
of a few comoving Mpc the amplitude of
the power spectrum reaches a maximum at the HRR, then decreases. This is true for case 1, which was used by Lidz et al. In
case 4, with the full Ly-$\alpha$ modeling, it looks like the maximum shifts to smaller values of the average ionization
fraction.

\section{Conclusions}

In this work, we have presented simulations of the $21\,\rm{cm}$ emission from the EoR. The first hydrodynamic simulations were run
in cosmological boxes of sizes $20\,h^{-1}\,\rm{Mpc}$ and $100\,h^{-1}\,\rm{Mpc}$ with $2 \times 256^3$ particles. Then, using
our adaptive Monte-Carlo code LICORICE, we ran, in post-treatment, radiative transfer simulations for the ionizing 
continuum. Finally 
 cosmological radiative transfer simulations in the Lyman-$\alpha$ line were performed.  The latter are necessary to compute the
neutral hydrogen spin temperature without making any assumption about the strength of the coupling. This approach has not been followed before.
 We model the sources of both ionizing continuum
and Lyman-$\alpha$  photons as stars, using a Salpeter IMF cut off at $100$ M$_\odot$. The $100\,h^{-1}\,\rm{Mpc}$ simulation
star formation history is calibrated on the $20\,h^{-1}\,\rm{Mpc}$. The photon escape fraction is calibrated independently
in both simulations to produce a Thomson scattering optical depth of CMB photons within $1\sigma$ of the WMAP3 value, and
a complete reionization at redshift $z\sim 6$. We have not included X-ray sources. If we had they would heat up the IGM and
reduce the contribution of absorption vs emission in the signal.

We first computed the evolution of a number of box-averaged quantities as functions of the average ionization 
fraction. We find that, with our source modeling, the average Lyman-$\alpha$ coupling coefficient $x_\alpha$ reaches
1 for small values of the ionization fraction (a few $10^{-3}$). However full coupling ($x_\alpha \gg 1$) is achieved
only for ionization fractions greater than $10^{-1}$. This is in line with expectations. Next we plotted the 
average spin temperature and differential brightness temperature.  It was shown that the results are very different
depending on how the average is computed~: directly on the spin$\backslash$brightness temperature of the particles, or from the
volume-averaged values of the gas kinetic temperature and coupling coefficients. The latter is often used, but the
former is more relevant. The direct average tends to erase the signal as regions in absorption have a much
stronger signal than regions in emission. Maps of the $21\,\rm{cm}$ emission confirm that the different assumptions made in 
computing the spin temperature lead to very different levels of the signal. The $T_S \gg T_{CMB}$ assumption is not relevant 
at all for
our choice of source modeling. It would be more relevant at not-too-early redshifts if we included X-ray sources.
 As expected the full coupling assumption, $x_\alpha = \infty$, is reasonable except in the
early phase of reionization. In this early phase, we illustrate how computing the local values of $x_\alpha$ produces
large fluctuations in the brightness temperature compared to using a redshift-dependent uniform value. 

Finally  the 3D dimensional power spectrum of the differential brightness temperature was plotted, with different
assumptions in computing the spin temperature, at various stages of reionization and for the two box sizes. The first
important (and expected) result is that including the possibility of a signal seen in absorption adds a large amount of
amplitude to the power spectrum. This obviously depends on the choice of the source model. Then, in the early phase of
reionization ($\langle x_H \rangle < 0.1$), using the local value of the Lyman-$\alpha$ coupling changes the spectrum
by transferring power from large scales to small scales. We reproduce the behaviour shown by Mellema et al. (\cite{Mellema06b}) and Lidz et al. 
(\cite{Lidz07b})~: a maximum of the amplitude at $\langle x_H \rangle=0.5$ for the scales around $10\,h^{-1}\,\rm{Mpc}$. 
However, when  absorption and the exact computation of the Lyman-$\alpha$ coupling are included, 
 this maximum is shifted to smaller ionization fractions.

As mentioned several times, the first modification to our model that will be investigated is a modification of the
source model. Including even a limited number of X-ray sources such as quasars would more efficiently preheat
the regions outside of the ionization front, reducing the strength of the absorption regime, or even turning it to
emission. This will be investigated  in a forthcoming paper.

Another interesting development of this work is to include the anisotropic effect of the peculiar velocities on the
$21\,\rm{cm}$ emission (Barkana \& Loeb \cite{Barkana05a}). Using the anisotropy, it is possible to separate the contribution
of the velocities to the $21\,\rm{cm}$ fluctuations from the other contributions and to derive constraints on the cosmological 
parameters. Since this effect is important mostly in the early reionization phase and on small scales, it is a logical
follow-up of including the full modeling of the Lyman-$\alpha$ coupling.

The implication of this work for the design of the future radio interferometers is simply to emphasize the
potential for the detection of the signal during the early phase of reionization when the fluctuations in the signal
are strong. Unfortunately, translating "early phase" into a numerical value for the redshift depends crucially on the
star formation history at these high redshifts, of which very little is known.
 From the value of the Thomson scattering optical depth, it is likely, however, that this early phase occurs at $z > 12$. 

\begin{acknowledgements}
This work was realized  in the context of the SKADS and HORIZON projects.
PDM acknowledges support from a SKADS post-doctoral fellowship.
The author thanks the anonymous referee for helpful comments and improvements.
\end{acknowledgements}

\end{document}